\documentclass[aps,prb,twocolumn,superscriptaddress,floatfix]{revtex4-2}

\usepackage{graphicx}
\usepackage{amsmath}
\usepackage{amssymb}
\usepackage{xcolor}
\usepackage{bm}
\usepackage{booktabs}
\usepackage{tabularx}
\usepackage{array}
\usepackage{siunitx}
\usepackage[colorlinks=true,citecolor=blue,linkcolor=blue,urlcolor=blue]{hyperref}

\newcommand{\RR}{\mathcal{R}}
\newcommand{\GG}{\mathcal{G}}
\newcommand{\HH}{\mathcal{H}}
\newcommand{\kq}{\mathbf{k}}
\newcommand{\Upi}{U_{\pi}}

\begin{document}

\title{Intrinsic $i$-wave altermagnetism in 2D graphene superlattices}

\author{Cuiju Yu}
\author{Jose L. Lado}
\affiliation{Department of Applied Physics, Aalto University, 02150 Espoo, Finland}

\begin{abstract}
Altermagnets feature unconventional
magnetism due to their momentum-dependent
spin splitting purely driven by magnetic order, for which
a variety of transition-metal-based $d$-wave altermagnets have been proposed.
However, carbon-based altermagnets remain largely unexplored, even though magnetism of graphene nanostructures has been widely demonstrated. 
Here, through a comprehensive symmetry analysis, we establish a symmetry-guided design principle for realizing higher-order $i$-wave altermagnetism in graphene antidot superlattices and demonstrate its emergence in both monolayer and bilayer graphene, particularly in the Bernal-stacked bilayer. By combining atomistic tight binding models and first principles methods, we show the appearance of an interaction-induced $i$-wave altermagnetic splitting, originating from the intrinsic magnetic instability of 2D graphene antidot superlattices. Our work establishes a strategy to engineer $i$-wave altermagnetism in a graphene platform, extending the materials landscape beyond conventional $d$-wave altermagnets and putting forward a carbon-based platform for higher-order altermagnetic spintronics.
\end{abstract}

\maketitle

\begin{figure*}[t!]
    \centering
    \includegraphics[width=.9\linewidth]{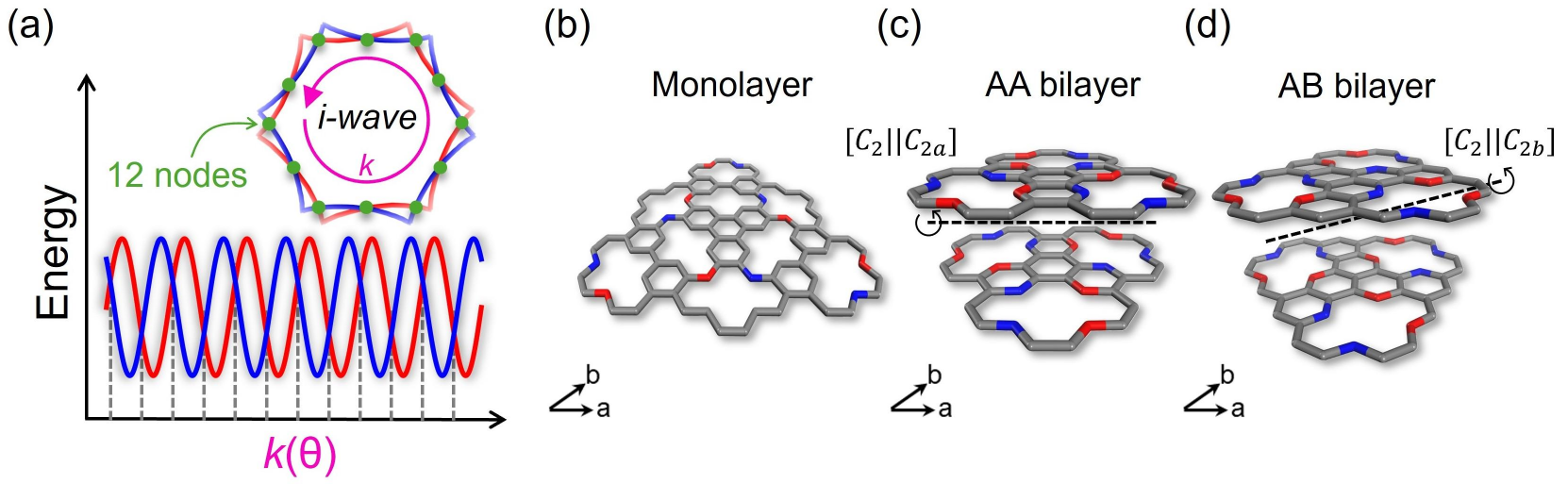}
    \caption{(a) Schematic representation of planar $i$-wave spin-momentum locking character protected by spin-point group symmetries across reciprocal space. Also shown is the spin-split band structure of the proposed $i$-wave altermagnet. Dashed gray lines drawn through the crossing points mark 12 nodes over the angular range from 0 to 2$\pi$, consistent with the $i$-wave altermagnetism. 
    (b-d) Schematic of the divacancy-induced $i$-wave altermagnetism in monolayer and bilayer regimes. Panel (b) shows a divacancy-engineered monolayer with spin point symmetry \(^{1}\bar{{6}}^{2}{m}^{2}{2}\) in 2D $k$-space, in the notation of \v{S}mejkal et al.\cite{vsmejkal2022beyond,vsmejkal2022emerging}. Panels (c,d) depict symmetrically arranged antidots in the AA-stacked (c) and AB-stacked (d) bilayers, giving rise to planar spin point group symmetries {\(^{1}{6}^{2}{2}^{2}{2}\) }(or {\(^{1}{6}^{2}{m}^{2}{m}\))} and  {\(^{1}\bar{{6}}^{2}{m}^{2}{2}\)}, respectively. 
    Here, red and blue short bars indicate local up and down magnetization.
    }
    \label{fig:1}
\end{figure*}

\section{Introduction}

Altermagnets (AMs) provide flexible 
platforms to explore unconventional magnetic order, simultaneously sharing properties of ferromagnets and antiferromagnets\cite{jungwirth2026symmetry,bai2024altermagnetism,song2025altermagnets,fender2025altermagnetism,guo2025spin}. 
AMs combine vanishing net magnetization and negligible stray field\cite{giil2024superconductor}, as in antiferromagnets with fully compensated magnetic moment, with spin-split band structures, as in ferromagnets\cite{krempasky2024altermagnetic}.
AMs hold great promise for realizing nontrivial topological states\cite{bai2026emergent,zou2025floquet,zhang2025altermagnetic,zhang2026topological}, and energy-efficient, highly integrated spintronic devices\cite{chen2025spin,tamang2025altermagnetism},
attracting considerable attention in condensed matter physics and spintronics.  AMs can be classified according to the symmetry of the spin-splitting in momentum space, leading to $d$-wave, $g$-wave, and $i$-wave types\cite{ezawa2025third}, similar to the classification of superconducting order parameters. Among these, $d$-wave altermagnets, typically realized in tetragonal rutile structures such as RuO$_2$\cite{smejkal2023chiral,chen2025spin}, have been extensively explored so far\cite{jiang2025metallic}. Such systems usually exhibit pronounced anisotropy of the effective masses along the $k_x$ and $k_y$ axes in momentum space, which translates into direction-dependent spin currents\cite{PhysRevLett.133.146602,liu2025realizing}. In such systems, ideal altermagnetic phases can be engineered through diverse structural-engineering strategies, for instance interlayer sliding\cite{zhu2025sliding}, Janus structures\cite{liu2025realizing} and twisting\cite{liu2024twisted}.  Interestingly, dual-switchable multiferroic altermagnets~\cite{gu2025ferroelectric,zhu2025two,fu2026two,duan2025antiferroelectric,peng2025ferroelastic} have emerged as a rapidly developing route to electrically and structurally control spin splitting.
Besides the intrinsic altermagnets, extrinsic altermagnets can be realized by applying external stimuli such as electric fields\cite{wang2024electric} and mechanical strain\cite{xu2025alterpiezoresponse}, further enriching the design routes and enhancing the tunability of altermagnetic states, for example in RuO$_2$-based spin-orbit torque devices\cite{li2025fully}. Regardless of specific systems, altermagnetism typically arises when opposite spin sublattices are related by spin space group symmetries such as rotation, mirror, screw, rather than by translation or inversion\cite{naka2025altermagnetic}.  

Existing proposals for achieving altermagnetism are largely confined to systems containing transition metals\cite{jiang2025metallic,liu2025realizing,lee2024broken,amin2024nanoscale,Mazin2023,camerano2025multiferroic}, and realizations on metal-free platforms are almost unexplored. Graphene-based systems  have risen as a flexible playground to engineer magnetism\cite{Yazyev2010}. In particular,
ferromagnetic states naturally arise in zigzag graphene nanoribbons\cite{PhysRevB.59.8271,PhysRevLett.100.047209,PhysRevLett.102.227205,Ruffieux2016}, giving rise to large magnetically-induced
exchange splittings\cite{Magda2014}. Single hydrogenation of graphene gives rise to local $S=1/2$ magnetic moments\cite{GonzlezHerrero2016}, providing
a highly tunable platform to engineer controllable magnetic superlattices by controlling hydrogenation with scanning tunneling microscopy. Furthermore, nanographene systems\cite{PhysRevLett.99.177204,su2019atomically,Mishra2021,PhysRevLett.124.177201,Mishra2019} provide a platform to 
create effectively zero and one-dimensional magnetic systems,
enabling the realization of paradigmatic models of quantum magnetism\cite{Sun2025,Zhao2025}, including topologically ordered states\cite{Mishra2021frac,Zhao2024}.
Furthermore, graphene multilayer systems have been shown to lead to magnetically ordered states stemming from
valley polarization, both in the case of twisted multilayers\cite{Serlin2020,Xie2021} and rhombohedral multilayers\cite{Lu2025,Han2024,Han2023}.
Despite the diverse realizations of graphene magnetism, metal-free altermagnetism has only recently been proposed in nanographene frameworks~\cite{ortiz2026organic} and H-engineered monolayer graphene~\cite{vina2025building}. Building on these advances, our design realizes a higher-order $i$-wave altermagnetism and further generalizes this symmetry class to bilayer graphene, particularly the previously unexplored Bernal-stacked configuration.
A higher-order $i$-wave state is also not detected in the same way as a $d$-wave one, since the linear responses used for $d$-wave altermagnets are blind to an $l=6$ form factor, and momentum-resolved probes such as the chiral magnon splitting take their place\cite{hoyer2025hematite,chiral2024mnte,sarkar2025magnon}.

Here we demonstrate the emergence of $i$-wave altermagnetism in graphene antidot superlattices, where magnetism stems purely from carbon. We propose a design rule for a 2D metal-free hexagonal lattice, achieving unconventional $i$-wave altermagnetism both in monolayer and bilayer graphene superlattices. We show that the crucial ingredient
is to design an antidot superstructure
with threefold rotational symmetry in which every mirror of the lattice exchanges the two spin sublattices, which accommodates $i$-wave altermagnetism. We demonstrate this strategy by
combining atomistic tight binding models with first-principles density functional theory methods, showing
the robustness of the $i$-wave altermagnetic state, in particular against lattice relaxations. 
Our manuscript is organized as follows. In Sec. 1, we provide an overview of recent advances in altermagnetism and nanographene systems. In Sec. 2, we detail general design principles in graphene antidot superlattice, followed by a mean-field Hubbard model validation and material realization via first principles simulations to demonstrate $i$-wave altermagnetism in graphene antidot superlattices.

\section{Graphene altermagnetic antidot superlattice}\label{sec:design}

Monolayer graphene has a non-magnetic ground state, although its electronic interactions put it at the verge of an electronic instability\cite{schuler2013optimal}.
Graphene multilayers, including aligned bilayer graphene, develop such an electronic instability\cite{Grushina2015,Velasco2012,freitag2012spontaneously,Weitz2010}.
The essence of our strategy is to leverage such electronic instability in a modified
graphene geometry, either in monolayer or bilayer, whose lattice symmetries give rise to an unconventionally altermagnetic $i$-wave order (Fig. \ref{fig:1}a).
This is realized by
a set of $C_3$ antidots arranged in a pattern whose mirror planes all exchange the two magnetic sublattices, leading to
an antidot graphene superlattice\cite{PhysRevLett.100.136804,Frst2009,furst2009density,PhysRevB.77.245431,Sandner2015}, as shown
in Fig. \ref{fig:1}b.
From the perspective of the spin symmetry group, the present model belongs at least to the $C_{3z}$ symmetry together with a spin-exchanging symmetry, which accommodates the presence of $i$-wave altermagnetism. The overall planar sixfold symmetry elements of the antidot lattice, realized through rotational or rotoinversion operations, produce an effective honeycomb network, essential for $i$-wave spin-momentum locking, as detailed below. It is worth noting that
there are no unpaired electrons in the structure, as each dot arises
from the removal of two sites in the original honeycomb lattice.

From the
experimental point of view, such systems can be constructed in two different ways.
A first option is a bottom-up synthesis of 2D nonplanar triple-porous nanographene with threefold symmetry on metal substrates, which has been demonstrated with on-surface reactions such as cyclodehydrogenation\cite{zuzak2019synthesis,xu2019surface,ajayakumar2022surface,qin2023synthesis}, yielding a structural framework closely resembling that of our proposed modified graphene.
A second strategy relies on effectively removing electronic $p_z$ orbitals in selected carbon sites by hydrogen adatoms,
as demonstrated using atomic manipulation with scanning tunneling microscope (STM)\cite{GonzlezHerrero2016,vina2025building}. Specifically, when a
single hydrogen adatom is added on top of a carbon site, the $\pi-$electron of carbon forms a covalent bond with H, leading to an effective removal of the $p_z$ from the
underlying low energy model\cite{PhysRevB.75.125408,PhysRevB.77.035427,PhysRevB.85.115405}. This fabrication strategy has the advantage that no chemical synthesis is needed, with the limitation that creating a bilayer structure can be more challenging.

Lieb's theorem\cite{PhysRevLett.62.1201} implies that total spin magnetization is strictly zero when the numbers of A- and B-sublattice sites are equal for the Hubbard model in a bipartite lattice. The intrinsic correlated antiferromagnetic (AFM) ground state in the antidot superlattice thus satisfies this constraint for achieving altermagnetism\cite{palacios2008vacancy}. Analysis of 20 2D centrosymmetric and non-centrosymmetric spin point groups (PG) identifies four candidates for planar $i$-wave altermagnetism:  $^16/^1m^2m^2m$, $^16^22^22$, $^16^2m^2m$ and $^1\bar{6}^2m^22$\cite{vsmejkal2022beyond,vsmejkal2022emerging}.  While a $C_6$ symmetry constraint can be readily realized in the monolayer limit through artificial divacancies, spin is intrinsically locked to the real-space orientation of antidots because of its AFM nature, thus yielding the $^26$ symmetry, instead of our desired $^16$. In this way, the first three candidates are excluded from realizing the $i$-wave character. Interestingly, the remaining case of $^1\bar{6}^2m^22$ makes $i$-wave altermagnetism plausible as it belongs to a reduced symmetry relative to the other three cases while remaining compatible with the AFM nature in antidot superlattices. As illustrated in Fig.~\ref{fig:1}b, a notable finding is that $^1\bar{6}^2m^22$ can be realized by deliberate engineering of P\(\bar{6}\)-symmetric antidots in a monolayer superlattice. This requires antidots to exhibit sixfold rotoinversion symmetry corresponding to PG $C_{3h}$ with identity operation in spin space ($^1\bar{6}$), while preserving a mirror plane and an in-plane twofold axis, each accompanied by a twofold rotation in spin space ($^2m$ and $^22$), so that the two opposite-spin sublattices are related by these operations rather than by translation or inversion. 
It should be noted that the highest symmetry retained by the spin arrangement is only $^1\bar{6}$, rather than $^26$.

Going beyond the monolayer limit, we generalize our strategy to bilayer graphene (BG). BG hosts interlayer AFM correlations due to Lieb's theorem, which stabilizes an antiparallel spin arrangement\cite{wang2013layer}. Taking advantage of the layer degree of freedom, we focus on two representative stacking forms, AA and AB (Bernal), to figure out how to realize intrinsic $i$-wave altermagnetism. Since two monolayers are perfectly aligned in the AA-stacked configuration, a planar $i$-wave pattern can be obtained by tailoring six pairs of antidots along hexagonal edges of graphene. In other words, the opposite spin lattices are connected by a $C_{2a}$ rotation, which maps the $C_3$-symmetric antidot pattern in the upper (lower) layer onto the lower (upper) layer (Fig.~\ref{fig:1}c), generating six evenly distributed petals with 12 nodal points (Fig.~\ref{fig:1}a).  Meanwhile, the intra- and inter-layer AFM correlations maintain antiparallel in-plane magnetic orders in the $C_6$ symmetric antidot lattice, thereby sustaining the planar spin-momentum locking of the spin PG $^16^22^22$ (Fig.~\ref{fig:1}c) and achieving the $i$-wave altermagnetism\cite{vsmejkal2022emerging,vsmejkal2022beyond}.
For AB stacking, it can be viewed as in-plane sliding of AA stacking. However, the resulting shift of rotation center completely breaks the whole symmetry when the same strategy as for AA stacking is applied. The loss of centrosymmetry makes spin-momentum locking difficult to realize by conventional means. From the symmetry analysis, the first three spin PG candidates are ruled out by lower $D_{3d}$ symmetry of pristine AB stacking, rather than $D_{6h}$ in AA. This leaves $^1\bar{6}^2m^22$ as allowed spin PG to realize $i$-wave altermagnetism. As shown in Fig.~\ref{fig:1}d, the antidot pairs are arranged to respect a sixfold rotoinversion axis in each layer, thereby satisfying $^1\bar{6}$ symmetry, whi le $^2m^22$ can be solely generated by in-plane $C_{2b}$ symmetry, thus the overall planar $^1\bar{6}^2m^22$ spin PG symmetry is successfully encoded into Bernal-stacked graphene superlattice. Beyond the spin point-group analysis, we also perform a rigorous and complete spin space group (SSG) analysis (see Appendix~\ref{app:ssg}).

\begin{figure}[t]
\centering
\includegraphics[width=\columnwidth]{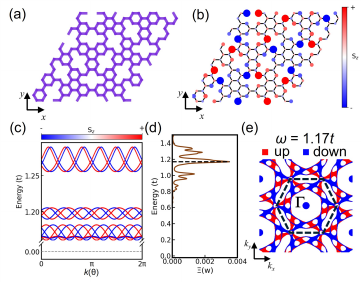}
\caption{(a) Top view of an antidot $6\sqrt{3}\times 6\sqrt{3}$ monolayer graphene superlattice. (b) Self-consistent spin density distributions. Here, red and blue solid circles represent spin up and spin down moments, respectively, and their sizes indicate the magnitudes of magnetic moments. (c) Spin-resolved band structure hosting spin-splitting character, in a path centered at the $\Gamma$ point of the Brillouin zone. Positive and negative values on the color bar denote spin-up and spin-down $S_z$ components, respectively.  The intralayer nearest-neighbor hopping is taken as the energy unit. The gray dashed line indicates the Fermi level. (d) Altermagnetic order density \(\Xi (\omega)\) versus energy. (e) Spin-resolved Fermi surface at $\omega=1.17t$ corresponding to the dashed line in (d).}
\label{fig:mon-tb}
\end{figure}

\subsection{Altermagnetism from self-consistent mean-field calculations}
\begin{figure*}[t!]
    \centering
    \includegraphics[width=1.\linewidth]{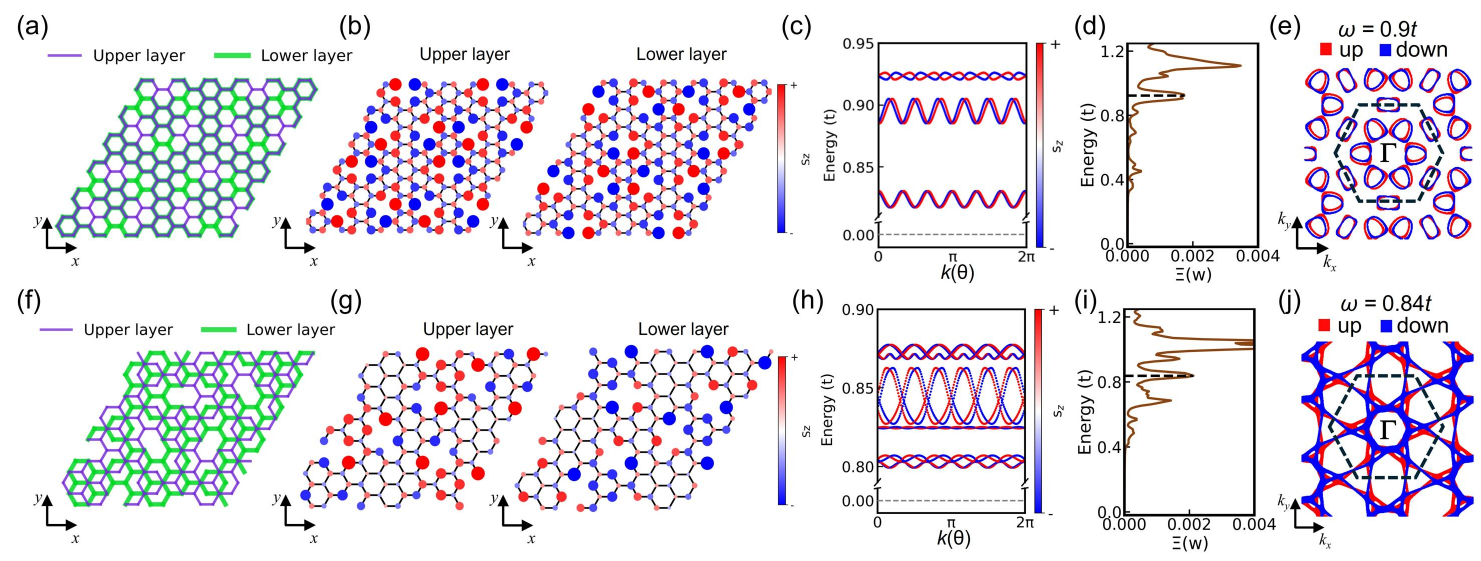}
    \caption{(a,f) Top-view schematics of antidot structures in $3\sqrt{3}\times 3\sqrt{3}$ AA-stacked, $5\sqrt{3}\times5\sqrt{3}$ Bernal-stacked graphene superlattices, respectively. (b,g) Corresponding self-consistent spin density distributions in top and bottom layers. Red and blue solid circles denote spin-up and spin-down magnetic moments, respectively, with their sizes proportional to the magnitudes of the actual local magnetic moments. (c,h) Spin-resolved band structures showing spin splitting for AA and AB stackings, respectively, in a path centered at the $\Gamma$ point of the first Brillouin zone. Here, the gray dashed lines indicate the Fermi level.  (d,i) Altermagnetic order density \(\Xi (\omega)\) as a function of energy for AA and AB stackings, respectively. Here the black dashed line marks the local maximum. (e,j) Spin-resolved Fermi surfaces centered at $\Gamma$ point for AA and AB stackings at $\omega=0.9t$ and $\omega=0.84t$, corresponding to the local maximum values in (d) and (i), respectively. It should be noted that, since the energy bands are exactly symmetric about the Fermi level, only the positive-energy bands are shown. }
    \label{fig:tb_bl}
\end{figure*}
To demonstrate the emergence of altermagnetism in those graphene superlattices,
we perform self-consistent calculations of an atomistic mean-field model.
We describe electronic structures
of the graphene antidot superlattice 
with a Hamiltonian of the form\cite{RevModPhys.81.109}
\begin{equation}
H = H_0 + H_U ,
\label{eq:hubbard_decomp}
\end{equation}
where $H_0$ is the non-interacting hopping term
\(H_0 = \sum_{ij ,\sigma} t_{ij}\, c_{i\sigma}^{\dagger} c_{j\sigma}\,\)
and $H_U$ is the local electrostatic interaction  \(H_U=U \sum_i c_{i\uparrow}^{\dagger} c_{i\uparrow} c_{i\downarrow}^{\dagger} c_{i\downarrow},\)
where $t_{ij}$ are the hoppings and $U$ is the on-site Hubbard interaction.  \(c_{i\sigma}^{\dagger} \) and \(c_{i\sigma} \) correspond to the creation and annihilation operators of spin \(\sigma\) on site $i$, respectively. The parameter \( t_{i j}\) represents hopping amplitude between different sites,
which include the intralayer hopping $t$ and the interlayer coupling $t_\perp$.  We set the intralayer nearest-neighbor hopping to  \(t\) = 2.7 eV and the interlayer nearest-neighbor hopping to \(t_\perp\) = 0.81 eV. 
We take a local Hubbard interaction of $U = 2t=5.4$ eV\cite{schuler2013optimal}. It is worth noting that for a pristine monolayer, such parameters give rise to a non-magnetic ground state as expected in pristine graphene. In contrast, an antidot superlattice will have an effective lower Fermi velocity, giving rise to a correlated state at the realistic interaction.
The interacting term H$_U$ is solved at the Hartree Fock mean-field level,
using a decoupling \(n_{i\uparrow} n_{i\downarrow} \approx \langle n_{i\uparrow} \rangle n_{i\downarrow} + \langle n_{i\downarrow} \rangle n_{i\uparrow} - \langle n_{i\uparrow} \rangle \langle n_{i\downarrow} \rangle\), yielding the self-consistent mean-field Hamiltonian $H_{\mathrm{MF}}$ parameterized by the local spin-resolved densities\cite{Pyqula}, which takes the form

\begin{equation}
    H_{\mathrm{MF}} = \sum_{ij,\sigma} t_{ij} \, c_{i\sigma}^{\dagger} c_{j\sigma}
    + \sum_{i,s,s'} \vec J_i \cdot \vec \sigma_{s,s'}
    c_{is}^{\dagger} c_{is'}
    \label{eq:meanfield}
\end{equation}
where $\vec J_i$ is the interaction-induced local exchange field, associated with
the emergence of a local magnetic order $\vec M_n = \sum_{s,s'} \vec \sigma_{s,s'}\langle c_{n,s}^{\dagger} c_{n,s'} \rangle$. Due to the bipartite nature of the system,
the mean-field ground state develops AFM order as given by Lieb's theorem\cite{PhysRevLett.62.1201}, and the self-consistent magnetization
is taken along the z-axis $\vec M_n = (0,0,M_n^z)$.

The appearance of altermagnetism can be directly quantified by the average altermagnetic order
defined as $
\Delta = \int d^2 \mathbf k \sum_\alpha \sqrt{\left(\epsilon_{\alpha,\uparrow}-\epsilon_{\alpha,\downarrow}\right )^2}
$
where $\epsilon_{\alpha,s}$ is the eigenvalue $\alpha$ for the spin channel $s$. 
The previous definition can be extended to the frequency range to quantify
which energies contribute the strongest to the altermagnetic order, by means of the altermagnetic spectral density defined as

\begin{equation}
\Xi (\omega) = \int d^2 \mathbf k \sum_\alpha \sqrt{\left(\epsilon_{\alpha,\uparrow}-\epsilon_{\alpha,\downarrow}\right )^2} \delta\left (\omega - \frac{\epsilon_{\alpha,\uparrow} + \epsilon_{\alpha,\downarrow}}{2}\right )
\end{equation}
where $\delta$ is the Dirac delta function. By definition we have $\Delta = \int  \Xi (\omega) d\omega$.

We first present the results for a monolayer graphene antidot superlattice (Fig.~\ref{fig:mon-tb}a). 
Self-consistent mean-field (SCMF) calculations give rise to a magnetic ground state,
for which the spin density distribution confirms that the ground state is AFM induced by the electronic instability (Fig.~\ref{fig:mon-tb}b). Subsequently, we analyze spin-resolved electronic bands by computing the $\langle S_z \rangle$ expectation value of each eigenstate.
We take a momentum space loop around the $\Gamma$ point 
in the Brillouin zone as $k(\theta) = k_0(\cos{\theta},\sin{\theta},0)$, where $k_0 = 0.3 |\mathbf b_1|$ with $|\mathbf b_1|$ the reciprocal lattice vector. A conventional $\Gamma$--K--M--$\Gamma$ path would show nothing, because the twelve nodal directions of an $l=6$ form factor are spaced by $30^\circ$ and the $\Gamma$--M and $\Gamma$--K directions of a hexagonal Brillouin zone are separated by exactly that angle, so both radial segments of that path are nodal. A loop of fixed $|k|$ instead sweeps the angle and exposes the sixfold sign alternation directly. 
As shown in Fig.~\ref{fig:mon-tb}c, the bands clearly show spin splitting in the reciprocal space, 
oscillating along the angular $k$-path coordinate $\theta$. 
Within the angular range $\theta \in [0,2\pi]$, twelve nodes (six pairs) of alternating spin splittings
are observed, consistent with the sixfold symmetry of $i$-wave altermagnetic order in Fig.~\ref{fig:1}a. By quantifying the altermagnetic order, we observe its density peaks at 1.17$t$ (Fig.~\ref{fig:mon-tb}d).
To further illustrate the emergence of $i$-wave order, we examine the Fermi surface at $\omega=1.17t$, as shown in Fig.~\ref{fig:mon-tb}e. The symmetrically distributed, petal-like spin-up and spin-down Fermi surfaces further confirm the presence of twelve nodal points in the Brillouin zone, consistent with split-ring features (Fig.~\ref{fig:mon-tb}c).

We now address the emergence of altermagnetism in bilayer superlattices.
We first consider the antidot superlattice in AA-stacked configuration depicted in Fig. \ref{fig:1}c. 
We perform SCMF calculations of the bilayer model (Fig.~\ref{fig:tb_bl}a), for which
the spin density distribution of upper and lower layers reveals the emergence of
a superlattice modulated antiferromagnetic order (Fig.~\ref{fig:tb_bl}b). Subsequently, we analyze the spin-resolved electronic band structure.
As shown in Fig.~\ref{fig:tb_bl}c, the bands exhibit a spin-dependent splitting in the reciprocal space, distributed along the angular $\theta$. 
For $\theta \in [0, 2\pi]$, six pairs of crossing splitting rings are observed, consistent with 12 nodal points in Fig.~\ref{fig:1}a.
As shown by the altermagnetic spectral density in Fig.~\ref{fig:tb_bl}d, a sizable spin splitting peak emerges at $\omega=0.9t$. 
The altermagnetic $i$-wave character can be directly observed on the Fermi surface (Fig.~\ref{fig:tb_bl}e). 
The spin-split Fermi surfaces alternate between hexagonal vertices in $k$-space and exhibit 12 nodes. 

We now consider the Bernal-stacked AB bilayer antidot superlattice.
The spin space symmetry of BG antidot superlattice belongs to the $^1\bar{6}^2m^22$, where the top view is shown in Fig.~\ref{fig:tb_bl}f.  The resulting mean-field
spin density distributions confirm the emergence of superlattice-modulated antiferromagnetic correlations between sublattices in the ground state
(Fig.~\ref{fig:tb_bl}g). From the band structure in a loop around the $\Gamma$ point, one clearly sees alternating spin-splitting loops, with 12 nodal points along the $k$-path in the range $\theta \in [0, 2\pi]$ in Fig.~\ref{fig:tb_bl}h. The frequency-resolved altermagnetic order density shown in Fig.~\ref{fig:tb_bl}i shows a sizable
altermagnetic ordering at the energy $\omega=0.84t$, as observed in Fig.~\ref{fig:tb_bl}h. The spin Fermi surface resembles two sets of distorted six-pointed stars, confirming the alternating-loop character of the electronic band and exhibiting 12 nodal points in first Brillouin zone (Fig.~\ref{fig:tb_bl}j). 
As a finite N\'eel temperature is not well defined for an ideal 2D antiferromagnet with negligible spin-orbit coupling (SOC), we instead extract the spin stiffness from the RPA magnon velocity, obtaining $\rho_s \approx 160$--$180~\mathrm{meV}$ for the monolayer and bilayers, six to seven times the room-temperature thermal energy, which indicates robust short-range antiferromagnetic correlations. The weak spin-orbit anisotropy of graphene cuts off the thermal magnon divergence and turns this stiffness into an estimated ordering temperature of about $245$~K for the monolayer and $225$~K for the bilayers (see Appendix~\ref{app:tloc}).

\subsection{Altermagnetism from first principles}
Density functional theory (DFT) calculations were performed using the Vienna ab initio Simulation Package (VASP)\cite{kresse1996efficient,kresse1996efficiency}. The interaction between electrons and ions was described using the projector-augmented-wave (PAW) method\cite{kresse1999ultrasoft,blochl1994projector}, while exchange-correlation effects were accounted for within the generalized gradient approximation (GGA)\cite{perdew1996generalized}. For structural optimization, the Perdew-Burke-Ernzerhof (PBE) functional is employed for computational efficiency. The energy cutoff was set to 600 eV for the plane wave basis set, with a dense 6$\times$6$\times$1 k-point mesh. The supercell structure was fully relaxed until the residual forces were below 0.001 eV/\AA, and the change of total energy was less than 10$^{-7}$ eV. The van der Waals interaction was accounted for using the DFT-D3 method for monolayer and bilayer graphene\cite{grimme2010consistent}. A vacuum layer of 17.0~\AA{} was introduced along the $c$-axis to prevent periodic interactions. Within both self-consistent and non-self-consistent calculations, we employed the meta-GGA SCAN functional\cite{sun2015strongly}, which is able to obtain a band gap closer to experiment. In pristine BG, the calculated band gap using SCAN functional is 8.3 meV, in reasonable agreement with experimentally reported values of 
about 8 or 2.5 meV\cite{veligura2012transport,freitag2012spontaneously}. Thermal and dynamical stability was evaluated by AIMD and phonon calculations (see Appendix~\ref{app:stability}).

\begin{figure}[t]
\centering
\includegraphics[width=\columnwidth]{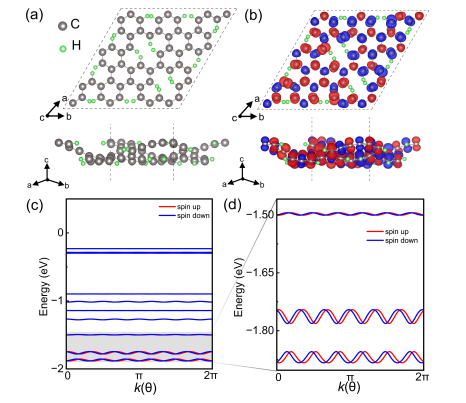}
\caption{(a) Schematic top and side views of divacancy-hydrogenated monolayer graphene of 6$\times$6 supercell. (b) Top and side views of spin density distribution in the ground state, where red and blue denote spin-up and spin-down densities. (c) Spin-resolved band structure along a circular path over the angular range from 0 to 2$\pi$. (d) Enlarged view of the spin-split bands in the highlighted region of (c).}
\label{fig:mon-dft}
\end{figure}
\begin{figure*}[!t]
    \centering
    \includegraphics[width=.9\linewidth]{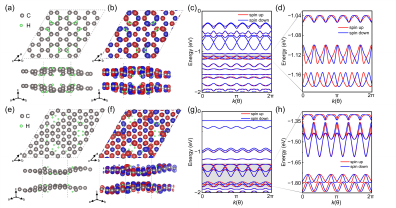}
    \caption{(a,e) Schematic top and side views of graphene superlattices,
    stemming from $3\sqrt{3}\times3\sqrt{3}$ AA-stacked and 5$\times$5 Bernal-stacked graphene supercells, respectively. (b,f) Top and side views of spin density distribution in ground states of AA and AB stackings, respectively, where red and blue shadows separately denote spin-up and spin-down densities. (c,g) Spin-resolved band structures within the angular range from 0 to 2$\pi$ 
    of a path around the $\Gamma$ point for AA and AB stackings, respectively. (d,h) Enlarged views of spin-split bands in gray highlighted regions of (c) and (g) for AA and AB stackings, respectively.}
    \label{fig:split_band} 
\end{figure*}

As a proof of concept of $i$-wave altermagnetism for the aforementioned antidot-engineered superlattices, we construct three representative material realizations, namely hydrogen-passivated vacancy superlattices in monolayer graphene, and in AA- and Bernal-stacked BG.
Firstly, we investigate structural, magnetic and electronic properties of the monolayer graphene antidot superlattice. Starting from $^1\bar{6}^2m^22$ symmetry in the unit cell with antidots, we find that the fully relaxed structure deviates from planarity and exhibits an out-of-plane distortion about 2.44 \AA, as shown in Fig.~\ref{fig:mon-dft}a. To further evaluate the magnetic ground state, total energies of nonmagnetic and AFM supercell configurations are analyzed. The AFM state lies 2.96 eV below the nonmagnetic state, establishing it as the ground state. The electronic band structure exhibits momentum-dependent spin splitting with 12 nodal points over the angular range $\theta \in [0,2\pi]$.
Despite hydrogen-induced spatial variations in the hopping and exchange fields, the out-of-plane buckling constitutes a symmetry-allowed structural distortion that preserves the characteristic symmetric 12-node $i$-wave pattern (Fig.~\ref{fig:1}a). The spin splitting of the buckled structure is comparable to that of the planar geometry (see Appendix~\ref{app:planar}). The resulting spin-momentum locking is a defining hallmark of $i$-wave altermagnetism.
It is worth noting that antidot-engineered supercell closely resembles porous graphene nanoribbons, where hydrogen-passivated vacancies are commonly adopted experimentally to stabilize the bare carbon edges. The same threefold-symmetric antidot arrangement in nonplanar triple-porous graphene nanostructures has been realized and characterized by STM and atomic force microscopy, which strongly supports the feasibility of our proposal~\cite{qin2023synthesis,xu2019surface}. 

We then turn to bilayer graphene antidot superlattice with AA- and Bernal-stacked configurations.
For the AA antidot superlattice in the minimal graphene supercell (Fig.~\ref{fig:split_band}a), the fully optimized structure develops buckled geometry along z axis, indicative of structural distortion driven by hydrogen-induced rehybridization from $sp^2$ to $sp^3$. To verify the magnetic ground state, three configurations were examined, namely nonmagnetic, interlayer ferromagnetic with intralayer AFM, and interlayer AFM with intralayer AFM. Their relative total energies are 1.27 eV, 0.109 eV, and 0 eV, respectively, demonstrating an AFM ground state (Fig.~\ref{fig:split_band}b). In addition to magnetic ground state, the electronic band structure exhibits a large band gap of 1.96 eV, confirming the insulating nature of the system, consistent with the gapped ground state of the model above. The spin-resolved band structure reveals alternating splitting from 0 to 2$\pi$ in reciprocal space (Fig.~\ref{fig:split_band}c,d). 
Notably, within the energy window from $-1.20$ to $-1.14$~eV, the intersecting spin-split loops form a sixfold pattern with 12 nodal points along a uniformly sampled circular path in the Brillouin zone, which serves a hallmark of $i$-wave altermagnetism. This result is consistent with the Hubbard-model prediction and further confirms that the symmetry-allowed $1.63$~\AA{} buckling within each individual graphene layer preserves the essential $i$-wave structure.
Despite this structural buckling reconstruction, the $i$-wave altermagnetic
order follows the results of the minimal atomistic model described previously. The buckling is expected to decrease in heterostructures encapsulated in hBN, as observed in other graphene devices\cite{PhysRevB.102.195404}.

We now address the Bernal-stacked antidot superlattice in the minimal graphene supercell as shown in Fig.~\ref{fig:split_band}e.
From a structural perspective, the Bernal-stacked antidot superlattice is energetically favored over the AA-stacked one by about 20 meV per superlattice unit cell. The AA-stacked configuration is nevertheless experimentally accessible, both as an extended phase in epitaxial bilayer graphene\cite{papagno2012two} and as locally AA-stacked regions of marginally twisted bilayer graphene\cite{ouyang2026structural}. The structure also undergoes out-of-plane buckling to relieve local strain, similar to the AA-superlattice case. From the total energy analysis and spin density distribution, the AFM configuration is energetically preferred to both the nonmagnetic and interlayer ferromagnetic states, by 1.308 and 0.109 eV, respectively (Fig.~\ref{fig:split_band}f).
We then move on to the electronic properties of divacancy-hydrogenated Bernal-stacked bilayer. As shown in Fig.~\ref{fig:split_band}g,h, within the energy window from $-1.87$ to $-1.35$~eV, one clearly sees six spin splitting loops and 12 nodal points along $k$-path, characteristic of altermagnetism with an $i$-wave component. 

In the shaded regions for AB stacking as shown in Fig.~\ref{fig:split_band}g, the lower two sets of alternating band spin splitting have complementary behaviors, showing a trend similar to that in the monolayer and AA-stacked cases. In addition, we examine the spin-split bands closer to the Fermi level (Fig. S6). 
The finite spin-split states emerge within 0.6~eV below the Fermi level, suggesting the $i$-wave character.
Moreover, fully relativistic calculations show that SOC induces only negligible magnetic anisotropy and band-structure modification as shown in Fig.~\ref{fig:soc_band}, leaving the characteristic $i$-wave altermagnetic spin splitting essentially unchanged (see Appendix~\ref{app:soc}). 
This momentum-dependent spin splitting 
makes the spin-resolved spectral weight strongly anisotropic, with the two spin channels inequivalent along selected momentum directions, e.g., $\pi/12$, where the spin splitting reaches its maximum in band structures in Fig.~\ref{fig:mon-dft}d and Fig.~\ref{fig:split_band}d,h, and equivalent along the nodal
directions. A charge-transport measurement is nevertheless not the appropriate probe of the net $i$-wave order. At charge neutrality these superlattices are magnetic insulators, and with the chemical potential in the gap the occupation factor is unity on every valence band and zero on every conduction band, so any Drude-type response reduces to the Brillouin-zone integral of a momentum derivative of a smooth periodic function, and vanishes identically at every order in the driving field. Independently of the filling, an $i$-wave form factor is an $l=6$ harmonic, so any Brillouin-zone-integrated response of rank five or lower is blind to it. The signature to target instead is the magnon spectrum.  In a compensated collinear altermagnet, the two chiral magnon branches acquire a momentum-dependent splitting that inherits the angular form factor of the electronic spin splitting. Consequently, the $i$-wave electronic structure is directly reflected in the magnon spectrum as a sixfold-periodic, sign-alternating splitting with twelve nodal directions in momentum space. This is accessible to magnon spin-transport measurements through the magnon spin Nernst and Seebeck effects in a collinear honeycomb geometry closely related to ours\cite{sarkar2025magnon}, while the electronic spin splitting that sets its form factor is directly accessible to spin-resolved angle-resolved photoemission spectroscopy (spin-ARPES) as demonstrated for altermagnetic MnTe\cite{krempasky2024altermagnetic} and CrSb\cite{yang2025three}, a technique that reaches states several eV below the Fermi level\cite{ngabonziza2025layer}, with micrometer-focused photoemission now resolving the bands of exfoliated graphene devices\cite{lisi2021observation}. Being nonrelativistic, the splitting requires no spin-orbit coupling, which suits a carbon platform (see Appendix~\ref{app:magnon}).

\section{Conclusion}
In summary, we showed the emergence of momentum-dependent spin-splitting associated with $i$-wave altermagnetism in graphene antidot superlattices, most notably in Bernal-stacked bilayer graphene.
Using mean-field Hubbard atomistic models and first principles methods based on density functional theory,
we showed that graphene superlattices enable the realization of $i$-wave altermagnetism in a pure carbon-based system. 
Our mechanism for altermagnetism relies solely on the intrinsic electronic interactions of graphene systems, combined with an engineered arrangement of missing
carbon sites realizing an antidot superlattice.
Altermagnetism in both monolayer and bilayer systems arises from a spatially modulated antiferromagnetic order triggered by the antidot superlattice
driven by electronic interactions, and we demonstrated that the $i$-wave order is robust against the presence of atomic relaxations in systems. 
Our results establish a strategy for engineering metal-free two-dimensional altermagnets in nanoengineered graphene systems, extending the material scope of altermagnetism beyond transition-metal compounds.

The Appendices provide additional calculations and benchmarks for graphene-based $i$-wave altermagnets, including the Hubbard-$U$ dependence, stability verification, projected band structures, SOC effects, magnetic anisotropy, and structural relaxation. It further presents the angular spin-splitting, spin space group and RPA spin response and spin stiffness analyses, local-moment formation temperature, the magnonic detection of the $i$-wave order, the supercell selection and the feasibility of AA-stacked bilayers. The relaxed structure files have been deposited in Zenodo (DOI 10.5281/zenodo.22658994).

\begin{acknowledgments}
We acknowledge the computational resources provided by the Aalto Science-IT project and the financial support from  InstituteQ,  
the Jane and Aatos Erkko Foundation, the Academy of Finland Projects Nos. 331342, 358088, and 349696, Finnish Ministry
of Education and Culture through the Quantum Doctoral Education Pilot Program (QDOC VN/3137/2024-OKM-4), the Research Council of Finland through the
Finnish Quantum Flagship project (358877, Aalto University),
the Finnish Center of Excellence in Quantum Materials QMAT (No. 374166), and the ERC Consolidator Grant ULTRATWISTROICS (Grant agreement no. 101170477).
We thank A. Fumega, L. Camerano and H. Wu for useful discussions.
\end{acknowledgments}

The data and computational results that support the findings of this study are available within the article and its Appendices. The relaxed structure files are deposited in Zenodo (DOI 10.5281/zenodo.22658994). Additional data, including relevant computational input files, are available from the corresponding authors upon reasonable request.

\appendix
\setcounter{topnumber}{1}
\setcounter{dbltopnumber}{1}
\renewcommand{\topfraction}{0.6}
\renewcommand{\dbltopfraction}{0.6}

\section{Tight-binding parameters and Hubbard interaction}\label{app:tb}

The hopping parameters follow standard tight-binding descriptions of graphene. The nearest-neighbor intralayer hopping is taken as $t=2.7~\mathrm{eV}$~\cite{reich2002tight}. The interlayer hopping is modeled using a distance-dependent Slater-Koster parametrization~\cite{trambly2012numerical}, as implemented in Pyqula~\cite{Pyqula}. For the interlayer spacing $d_z=3.0~\text{\AA}$, we use
\(t_\perp \approx 0.3t \approx 0.81~\mathrm{eV}\)
consistent with an exponential interpolation from $t_\perp(d_z=3.35~\text{\AA})\approx0.39~\mathrm{eV}$.

For the on-site Hubbard interaction, we use $U=2t=5.4~\mathrm{eV}$, comparable to the effective value $U_{\mathrm{eff}}\approx1.6t$ reported for graphene when nonlocal Coulomb interactions are included~\cite{schuler2013optimal}.

\section{Hubbard-$U$ dependence and second-nearest-neighbor hopping dependence}\label{app:hubbard}

\begin{figure}[t]
    \centering
    \includegraphics[width=\columnwidth]{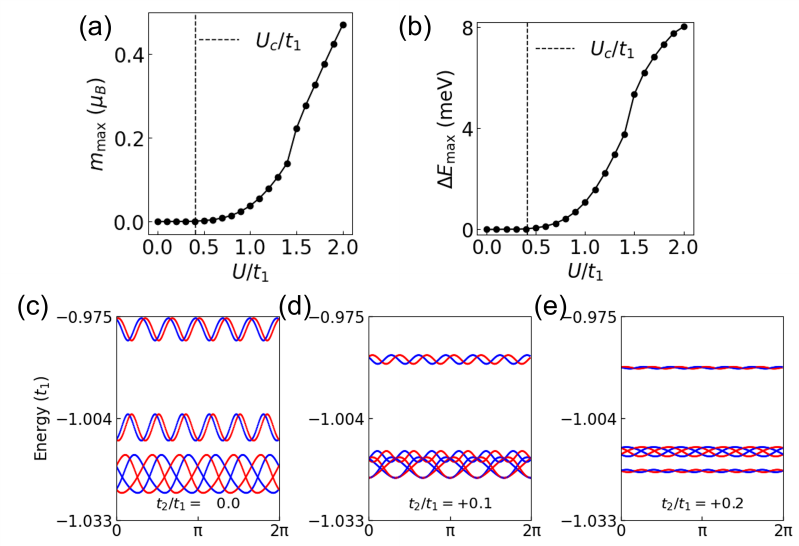}
    \caption{ For graphene monolayer superlattice, (a) maximum local magnetic moment per C site, $m_{\mathrm{max}}$, as a function of $U/t_1$. (b) Maximum spin-splitting energy, $\Delta E_{\mathrm{max}}$, as a function of $U/t_1$, evaluated along the same circular $\Gamma$-centered path. (c--e) Spin-resolved band structures for $t_2/t_1=0.0$, $0.1$, and $0.2$, respectively. Energies are given in units of $t_1$.}
    \label{fig:U_SCAN_tb}
\end{figure}

\begin{figure}[t]
    \centering
    \includegraphics[width=\columnwidth]{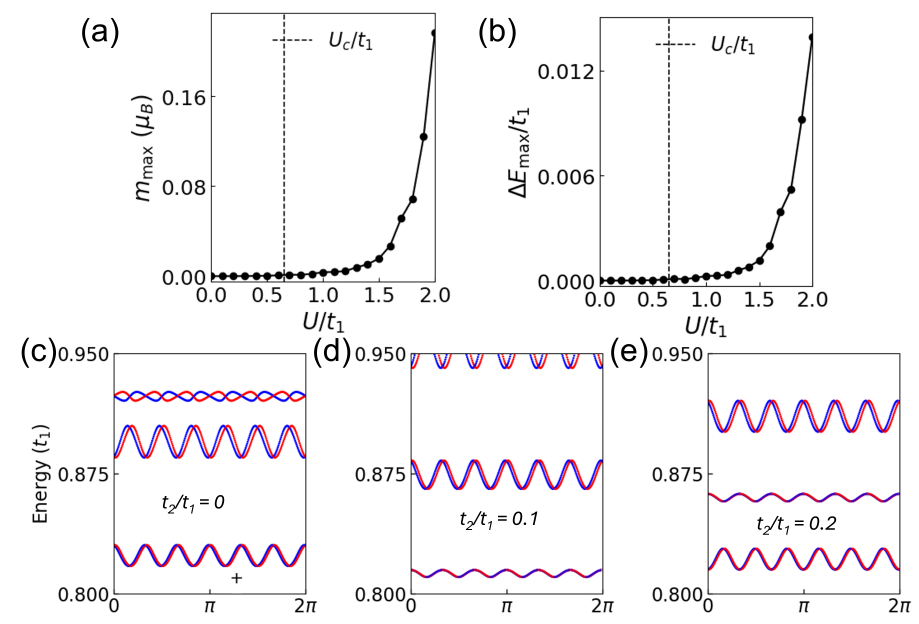}
    \caption{For AA bilayer graphene superlattice, (a) maximum local magnetic moment per C site, $m_{\mathrm{max}}$, as a function of $U/t_1$. (b) Maximum spin-splitting energy as a function of $U/t_1$. (c--e) Spin-resolved band structures for $t_2/t_1=0.0$, $0.1$, and $0.2$, respectively. }
    \label{fig:aa-scan}
\end{figure}

\begin{figure}[t]
    \centering
    \includegraphics[width=\columnwidth]{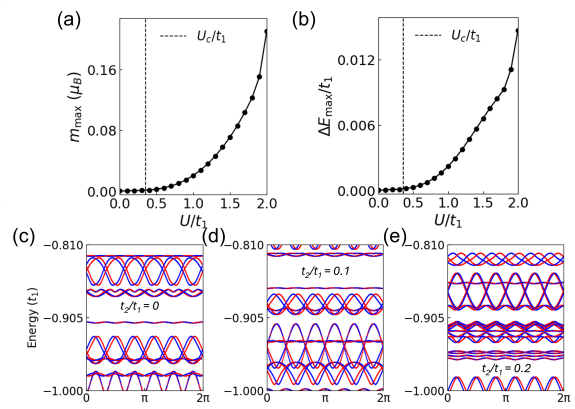}
    \caption{For AB bilayer graphene superlattice, (a) maximum local magnetic moment per C site, $m_{\mathrm{max}}$, as a function of $U/t_1$. (b) Maximum spin-splitting energy as a function of $U/t_1$. (c--e) Spin-resolved band structures for $t_2/t_1=0.0$, $0.1$, and $0.2$, respectively. }
    \label{fig:ab-scan}
\end{figure}

In this section $t_1$ denotes the nearest-neighbor intralayer hopping, written as $t$ elsewhere, and $t_2$ the second-nearest-neighbor hopping. We have scanned the Hubbard interaction over $U/t_1=0$--$2$ and examined its effect on the local magnetic moment per C site (in $\mu_B$) and on the spin-splitting energy for the three antidot superlattices. Both the maximum local magnetic moment and the spin-splitting energy generally increase with increasing Hubbard $U$. The critical interaction $U_c$ is defined by the onset of a stable self-consistent magnetic solution with a finite local moment. As shown in Fig.~\ref{fig:U_SCAN_tb}a,b, the $U$ scan places the onset at $U_c/t_1 \approx 0.41$ (dashed line), marking the transition from the nonmagnetic to the magnetic phase. Similarly, Figs.~\ref{fig:aa-scan}a,b and \ref{fig:ab-scan}a,b give $U_c/t_1 \approx 0.65$ for the AA bilayer and $U_c/t_1 \approx 0.35$ for the AB bilayer, respectively. In all three systems the moment stays small over a wide interval above the onset and grows rapidly only well above it.

We have also included the second-nearest-neighbor hopping $t_2$ in the model. For $t_2/t_1>0$ the $i$-wave altermagnetic pattern is insensitive to $t_2$, while the positions of the alternating spin-split loops shift, as seen in Fig.~\ref{fig:U_SCAN_tb}c-e, Fig.~\ref{fig:aa-scan}c-e and Fig.~\ref{fig:ab-scan}c-e. The spin splitting persists upon inclusion of $t_2$, with its magnitude evolving as $t_2$ increases and the local magnetic moment decreasing slightly. Therefore, we keep only $t_1$ in the minimal model, which captures the essential features of the $i$-wave altermagnetism.

The maximum spin splitting obtained from the tight-binding model and from density functional theory (DFT) is compared in Table~\ref{tab:spin_splitting}.

We have also compared the local magnetic moments obtained from the two approaches. The tight-binding model gives $0.247$, $0.152$ and $0.123~\mu_B$ per C site for the monolayer, the AA-stacked bilayer and the AB-stacked bilayer, against $0.186$, $0.143$ and $0.154~\mu_B$ per C site from the DFT calculations. The model and the DFT calculations use the same supercell only for the AA bilayer, so this comparison is qualitative rather than cell-by-cell. Within that caveat the two approaches yield local moments of the same order of magnitude in all three systems, showing that the model and the DFT description are consistent.

\begin{table}[t]
\centering
\caption{Maximum spin splitting obtained from the model and from first-principles calculations for three representative systems. The model values are maxima over the full Brillouin zone.}
\label{tab:spin_splitting}
\begin{tabular}{lcc}
\hline
System & $\Delta E_{\mathrm{max}}^{\mathrm{model}}$ (meV) & $\Delta E_{\mathrm{max}}^{\mathrm{DFT}}$ (meV)  \\
\hline
Monolayer  & 34 & 33  \\
AA-stacked  & 74 & 78   \\
AB-stacked & 50 & 45  \\
\hline
\end{tabular}
\end{table}

\section{Stability verification}\label{app:stability}

\begin{figure}[t]
    \centering
    \includegraphics[width=\columnwidth]{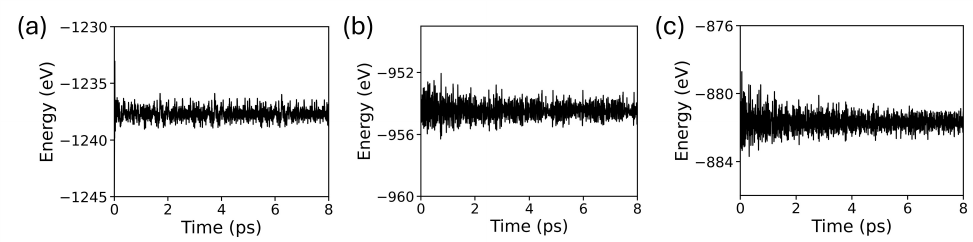}
    \caption{Total energy evolution during ab initio molecular dynamics (AIMD) simulations at 300~K over 8~ps for (a) the monolayer, (b) the AA-stacked bilayer and (c) the AB-stacked bilayer.} 
    \label{fig:aimd}
\end{figure}

\begin{figure*}[t]
    \centering
    \includegraphics[width=.9\linewidth]{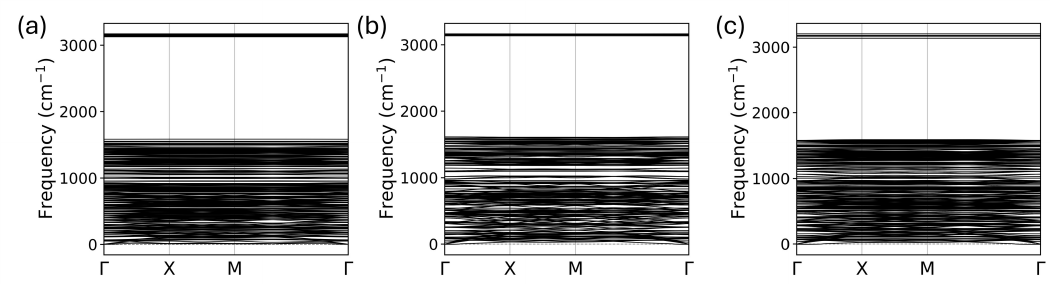}
    \caption{Phonon dispersions of the (a) monolayer, (b) AA-stacked bilayer, and (c) AB-stacked bilayer structures along the high-symmetry path.}
    \label{fig:phono}
\end{figure*}

We examined the thermal and dynamical stability of the three structures through ab initio molecular dynamics (AIMD) simulations at 300~K and phonon calculations, respectively. During the 8~ps of simulation the structures remain intact, without bond breaking or noticeable structural reconstruction (Fig.~\ref{fig:aimd}). The calculated phonon spectra show no imaginary modes throughout the Brillouin zone (Fig.~\ref{fig:phono}). The three systems are therefore stable against both finite-temperature fluctuations and small lattice perturbations.

\section{Spin-resolved band structures along different $k$-paths}\label{app:kpaths}

\begin{figure}[t]
    \centering
    \includegraphics[width=\columnwidth]{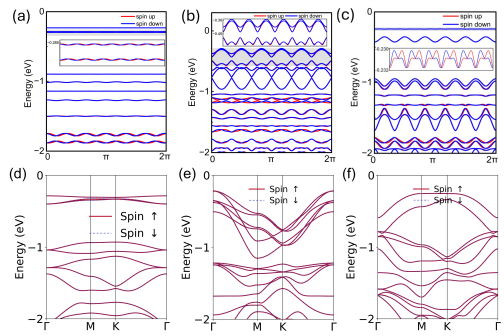}
    \caption{Spin-resolved band structures along (a--c) a circular $k$-path and (d--f) the conventional linear $k$-path, with enlarged views of the gray-shaded regions near the Fermi level, for (a,d) the monolayer graphene superlattice, (b,e) the AA bilayer and (c,f) the AB bilayer. The insets show the first Brillouin zone and the selected momentum-space paths. }
    \label{fig:split_bands_VBM}
\end{figure}

Fig.~\ref{fig:split_bands_VBM}a--c shows the first-principles spin-resolved bands on a circular path whose radius is chosen so that the states within 0.6 eV of the Fermi level carry a resolvable $i$-wave splitting. In the first-principles band structures the Fermi level is placed at the valence-band maximum, so this depth is measured from the band edge.
We also calculated the band structures along the conventional high-symmetry $k$-path, as shown in Fig.~\ref{fig:split_bands_VBM}d--f. The twelve nodal directions of an $l=6$ form factor are spaced by $30^\circ$, and in a hexagonal Brillouin zone the $\Gamma$--M and $\Gamma$--K directions are separated by exactly that angle, so both of them are nodal. Both radial segments of the conventional path therefore carry no altermagnetic splitting by symmetry, and only the zone-boundary segment samples off-nodal directions. This is why the bands are displayed on a circular path of fixed $|k|$ in panels a--c, which sweeps the angle and exposes the sixfold sign alternation directly.
In addition, varying the antidot size or spacing in graphene can tune the defect-state hybridization, magnetic moment, and band gap, and may shift the spin-split features closer to the Fermi level~\cite{furst2009density}.

We also quantify the doping requirement by the chemical-potential shift from charge neutrality needed to reach the spin-split states. In the tight-binding model, splittings of \(\sim10\) meV occur at \(0.35t\) and \(0.65t\) for the AA and AB bilayers, respectively, while the monolayer requires a larger shift of \(0.83t\) to reach \(8\) meV. DFT shows the same trend, with weak near Fermi level splittings of 1.6 to 8.1 meV and larger splittings of 16 to 18 meV at deeper energies. Because such large shifts may be difficult to achieve by gating, spin-ARPES provides an alternative probe of the deeper spin-split states (see Appendix~\ref{app:magnon}). A doping of this size would in any case quench the altermagnetic moment itself. The moments here form on the half-filled midgap states of the antidot lattice, which is the same mechanism that produces defect-induced magnetism in graphene~\cite{PhysRevB.75.125408,palacios2008vacancy,Yazyev2010}, and charging those states removes the local moment that the splitting is built on~\cite{PhysRevLett.99.177204}.

\section{Orbital-projected band structures}\label{app:pband}

\begin{figure}[t]
    \centering
    \includegraphics[width=\columnwidth]{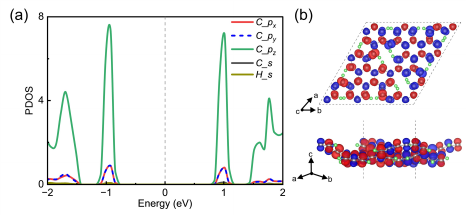}
    \caption{ (a) Orbital-projected density of states of the monolayer antidot superlattice. (b) Top and side views of spin density distribution in the ground state, where red and blue denote spin-up and spin-down densities.
}
    \label{fig:pband-mon}
\end{figure}

Using monolayer graphene as an example, we further identify the atomic orbitals contributing to the split bands. The projected-band analysis shows that these bands are dominated by C \(p_z\) orbitals, with minor contributions from C \(p_x\) and \(p_y\) orbitals, while the contribution from C \(s\) orbitals is negligible (Fig.~\ref{fig:pband-mon}a). As shown in Fig.~\ref{fig:pband-mon}b, these states are distributed across the lattice rather than localized at the antidot edges.

\section{Spin-orbit coupling and magnetic anisotropy}\label{app:soc}
To evaluate the SOC induced by the structural distortion, we performed fully relativistic calculations including SOC for the fully relaxed geometries. The magnetic anisotropy energy between the N\'eel-vector orientations along [001] and [100] is 0.017~meV per C site for the monolayer and 0.006~meV per C site for the bilayers, a negligible SOC-induced magnetic anisotropy. The intrinsic spin-orbit coupling of flat graphene is itself only $24~\mu$eV and, as first-principles calculations show, comes mainly from the $d$ orbitals~\cite{gmitra2009band}, so the anisotropies computed here for the buckled superlattices are of the same order.

\begin{figure}[t]
    \centering
    \includegraphics[width=\columnwidth]{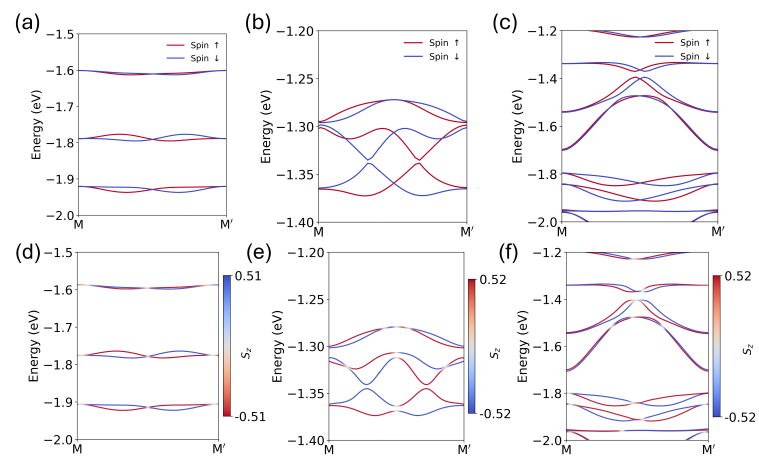}
    \caption{Spin-resolved band structures of the graphene superlattice calculated along the conventional high-symmetry $k$-path without SOC (a--c) and with SOC (d--f) for the monolayer, AA bilayer, and AB bilayer, respectively. }
    \label{fig:soc_band}
\end{figure}

As displayed in Fig.~\ref{fig:soc_band}, the band structure with SOC remains nearly indistinguishable from that without SOC, and the characteristic momentum-dependent spin splitting is preserved. Therefore, although SOC formally reduces the nonrelativistic spin-space symmetry, the deformation-induced SOC is a weak relativistic perturbation, three orders of magnitude below the exchange-driven altermagnetic splitting, and does not destabilize the $i$-wave altermagnetism.

\section{Comparison of planar and nonplanar monolayer band structures}\label{app:planar}

\begin{figure}[t]
    \centering
    \includegraphics[width=\columnwidth]{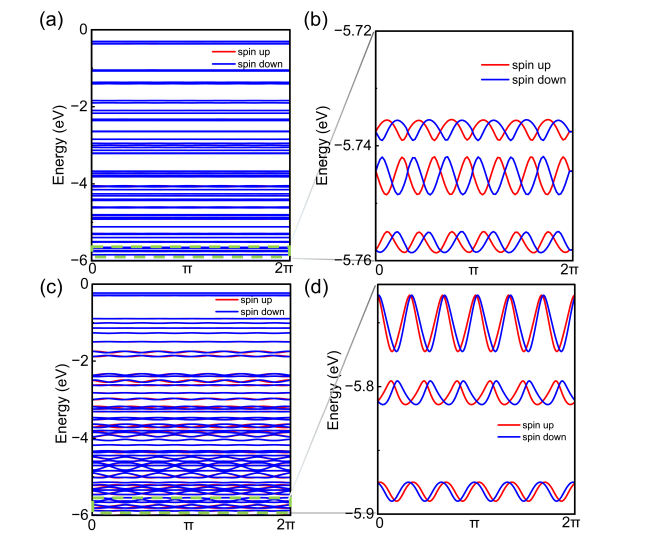}
    \caption{Band structures of monolayer graphene for (a,b) the planar structure, and (c,d) the fully relaxed nonplanar structure. Panels (b) and (d) show enlarged views of the regions highlighted by the green boxes in panels (a) and (c), respectively.}
    \label{fig:flatband}
\end{figure}

We relaxed the monolayer superlattice under a planar constraint, optimizing the in-plane atomic coordinates while holding the out-of-plane positions fixed, and computed its band structure (Fig.~\ref{fig:flatband}a,b). The $i$-wave altermagnetic pattern of the planar structure is unchanged and consistent with the tight-binding model, and its spin splitting is comparable to that of the fully relaxed geometry (Fig.~\ref{fig:flatband}c,d). The structural buckling is therefore a symmetry-allowed distortion and not a prerequisite for altermagnetism.

\section{Angular dependence of the spin splitting}\label{app:angular}

We analyze the signed splitting of the band $n$ closest to the Fermi level, defined as
\begin{equation}
\Delta E_n(k_r,\theta) = E_{n\uparrow}(k_r,\theta)-E_{n\downarrow}(k_r,\theta),
\end{equation}
and decompose it as
\begin{equation}
\Delta E_n(k_r,\theta)=
a_0+\sum_{m=2,4,6}
\left[a_m\cos(m\theta)+b_m\sin(m\theta)\right],
\end{equation}

where $A_m=\sqrt{a_m^2+b_m^2}$ is the amplitude of the $m$-th harmonic. This selection rule is exact. With real hoppings and negligible spin-orbit coupling the splitting satisfies $\Delta E_n(\theta+180^\circ)=\Delta E_n(\theta)$, while the $C_3$ symmetry of the antidot lattice imposes $\Delta E_n(\theta+120^\circ)=\Delta E_n(\theta)$. Taken together these give a $60^\circ$ period, so every harmonic below $m=6$ vanishes identically rather than merely being small.

\begin{figure}[t]
    \centering
    \includegraphics[width=\columnwidth]{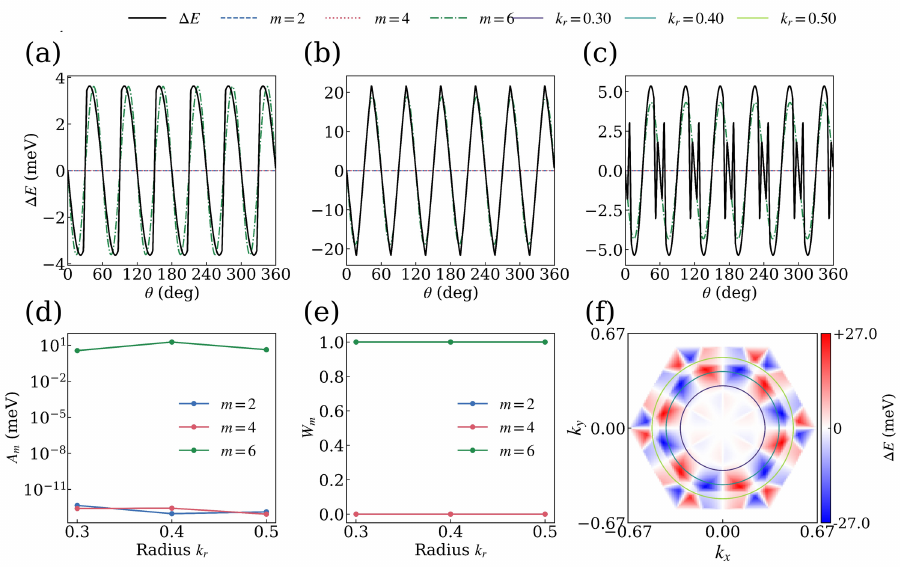}
    \caption{(a--c) Angular dependence of
$\Delta E_n(k_r,\theta)=E_{n\uparrow}(k_r,\theta)-E_{n\downarrow}(k_r,\theta)$
for the fixed band closest to the Fermi level in monolayer graphene, evaluated at
$k_r=0.30$, $0.40$, and $0.50$, respectively. The black curves show the calculated splitting, while the dashed, dotted, and dash-dotted curves denote the $m=2$, $m=4$, and $m=6$ harmonic components. 
(d,e) Radial dependence of the harmonic amplitudes $A_m$ and normalized weights $W_m$, demonstrating that the $m=2$ and $m=4$ contributions are negligible compared with the dominant $m=6$ component at all examined radii.
(f) Two-dimensional map of the signed spin splitting $\Delta E_n(\mathbf{k})$ in the first Brillouin zone. The colored circles indicate the momentum-space contours used in panels (a--e). The alternating sign pattern and robust $m=6$ dominance across different radii establish the characteristic $i$-wave spin-splitting symmetry.}
    \label{fig:i-wave_component}
\end{figure}

As shown in Fig.~\ref{fig:i-wave_component}a-c, the $m=2$ and $m=4$ components vanish to numerical precision, staying below $10^{-11}$~meV (Fig.~\ref{fig:i-wave_component}d), whereas the $m=6$ component dominates at every radius examined ($k_r = 0.3$, $0.4$, $0.5$). The vanishing angular average $a_0$ further excludes a uniform ferromagnetic-like splitting. The signed splitting also satisfies
\begin{equation}
\Delta E_n(\theta+\pi/6)\simeq-\Delta E_n(\theta),
\end{equation}
resulting in the twelve sign-changing sectors of an $i$-wave spin texture, and both the harmonic amplitudes and the normalized weights show that the $m=6$ component dominates (Fig.~\ref{fig:i-wave_component}d,e).

Fig.~\ref{fig:i-wave_component}f shows a two-dimensional map of $E_{n\uparrow}(\mathbf{k})-E_{n\downarrow}(\mathbf{k})$ across the first Brillouin zone, from which the angular profiles at $k_r = 0.3$, $0.4$ and $0.5$ are extracted. The same sign pattern and dominant $m=6$ component persist at all radii, so the result is not specific to a single momentum-space contour.

\begin{figure}[t]
    \centering
    \includegraphics[width=\columnwidth]{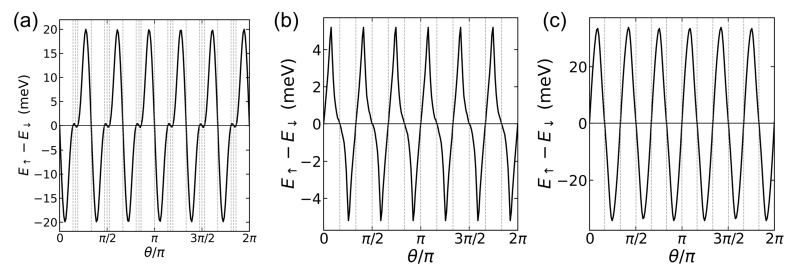}
    \caption{Angular spin splitting $\Delta E(\theta)=E_{\uparrow}-E_{\downarrow}$ for the fully relaxed
    (a) monolayer, (b) AA-stacked bilayer, and (c) AB-stacked bilayer structures.
    The gray dashed lines mark the 12 sign-changing nodal crossings. Here, the split band is chosen from the occupied state near the Fermi level.}
    \label{fig:angular_nodes_splitting}
\end{figure}

To quantify the effect of buckling on the nodal degeneracies, we evaluated the angular spin splitting
$\Delta E(\theta)=E_{\uparrow}(\theta)-E_{\downarrow}(\theta)$
along a closed momentum-space path for the fully relaxed structures.
As shown in Fig.~\ref{fig:angular_nodes_splitting}, $\Delta E(\theta)$ changes sign through 12 nodal directions in the monolayer, AA-stacked bilayer, and AB-stacked bilayer, with the interpolated splitting vanishing at each crossing within numerical resolution.
Buckling therefore modifies the magnitude and angular distribution of the spin splitting rather than lifting the nodal degeneracies.

\section{Spin space group analysis}\label{app:ssg}
We provide the explicit spin-space-group (SSG) analysis for the monolayer and the AA- and AB-stacked bilayers.
 
\subsection{ General decomposition}

For a compensated collinear magnet, the SSG part can be written as
\begin{equation}
    \mathcal{R}_{s} = \left[ E \parallel H \right] + \left[ U_{\pi} \parallel A H \right]
    \label{eq:gen}
\end{equation}
where $H$ is a halving subgroup of the parent (crystallographic) group $\GG$ ($[\GG{:}H]=2$) that preserves each spin sublattice, while $A$ ($\GG\setminus H = A\,H$) exchanges the two opposite-spin sublattices, combined with the spin-space $\pi$ rotation $\Upi$ that flips the spins. The primitive translations $[E\Vert E\,|\,\mathbf{a}_i]$ are implicitly included. Equation~\eqref{eq:gen} is a Type-III SSG. It describes an altermagnet, rather than a conventional antiferromagnet, provided that the spin-flip coset contains neither a pure translation nor the spatial inversion, so that the opposite-spin sublattices are linked by a rotation, mirror or screw operation~\cite{le2026catalog,naka2025altermagnetic}.

The spin point groups quoted in Sec.~\ref{sec:design} are planar spin point groups,
not the point-group parts of the three-dimensional SSG decomposition in
Eq.~\eqref{eq:gen}.  The two are related by restricting each SSG to its
action on the $k_x$--$k_y$ plane, and impose the same constraints on the
in-plane spin-splitting form factor and hence the same $i$-wave character. In
the decomposition below, the superscript-1 operations generate the
spin-preserving subgroup $H$ and the superscript-2 operations the
spin-exchanging coset $\GG\setminus H$.

\subsection{Monolayer}

For the ideal planar monolayer, the spin point group is $^{1}\bar{6}^{\,2}m^{2}2$ (crystal part $D_{3h}$, which contains no spatial inversion), one of the four planar spin point groups that host $i$-wave altermagnetism\cite{vsmejkal2022beyond,vsmejkal2022emerging}. The superscripts record which operations are paired with the spin flip. The sixfold rotoinversion $\bar{6}$, and with it the threefold rotation $C_{3z}$ and the horizontal mirror $\sigma_h$, carry the superscript 1 and preserve each spin sublattice, whereas the three vertical mirrors $\sigma_v$ (planes containing the $z$ axis) and the three in-plane twofold axes $C_2'$ carry the superscript 2 and exchange the two opposite-spin sublattices. In the fully relaxed monolayer the structure deviates from planarity (out-of-plane distortion $\sim2.44$\,\AA) and the space group is lowered to $P31m$ ($C_{3v}$). Buckling removes $\sigma_h$, and with it $\bar{6}$ and $C_2'$, but the two generators on which the splitting rests survive, namely
\begin{equation}\label{eq:mono}
  \begin{gathered}
  \bigl[\,E\,\Vert\,C_{3z}\,\bigr]\ \text{(spin-preserving)},\\
  \bigl[\,U_\pi\,\Vert\,\sigma_v\,\bigr]\ \text{(spin-transposing)},
  \end{gathered}
\end{equation}
giving
\begin{equation}\label{eq:mono-ssg}
  \RR_s^{\mathrm{ML}}=\bigl[\,E\,\Vert\,P3\,\bigr]+\bigl[\,\Upi\,\Vert\,\sigma_v P3\,\bigr].
\end{equation}
Here $C_{3z}$ preserves each spin sublattice, whereas $\sigma_v$ (a vertical mirror plane containing the $z$ axis) exchanges the two opposite-spin sublattices and
therefore imposes
\[
\Delta E(\sigma_v\mathbf{k})
=
-\Delta E(\mathbf{k}).
\]

\subsection{AA- and AB-stacked bilayers}

For the AA-stacked bilayer, the space group is $P\bar{3}m1$, which is preserved both before and after buckling.
The spin-preserving subgroup is
\[
H_{\mathrm{AA}}=P\bar{3},
\]
while the remaining operations form the spin-exchanging coset
\[
P\bar{3}m1 = P\bar{3}\cup C_{2a}P\bar{3}.
\]
The corresponding spin space group is
\begin{equation}\label{eq:bil}
  \RR_s^{\mathrm{AA}} \;=\; \bigl[\,E\,\Vert\,P\bar{3}\,\bigr] \;+\; \bigl[\,U_\pi\,\Vert\,C_{2a}\,P\bar{3}\,\bigr], 
\end{equation}
with generators
\[
[E\Vert IC_{3z}],\qquad
[U_{\pi}\parallel C_{2a}].
\]
Here, the inversion centered on the interlayer midplane exchanges the two layers
while preserving the magnetic sublattices, whereas $C_{2a}$ exchanges the
opposite-spin sublattices and therefore imposes
\[
\Delta E(C_{2a}\mathbf{k})
=
-\Delta E(\mathbf{k}).
\]

In the AB-stacked bilayer, the Bernal shift reduces the symmetry to the space group $P321$, which is preserved both before and after buckling.
The spin-preserving subgroup becomes
\[
H_{\mathrm{AB}}=P3,
\]
and the spin-exchanging coset is
\[
P321 = P3 \cup C_{2b}P3 .
\]
Accordingly,
\begin{equation}\label{eq:bil2}
  \RR_s^{\mathrm{AB}} \;=\; \bigl[\,E\,\Vert\,P3\,\bigr] \;+\; \bigl[\,U_\pi\,\Vert\,C_{2b}\,P3\,\bigr],
\end{equation}
with generators
\[
[E\parallel C_{3z}],\qquad
[U_{\pi}\parallel C_{2b}].
\]
The spin-exchanging $C_{2b}$ operation similarly gives
\[
\Delta E(C_{2b}\mathbf{k})
=
-\Delta E(\mathbf{k}).
\]

Both bilayers thus share the same spin-symmetry structure, in which $C_{3z}$ preserves the magnetic sublattices, whereas an in-plane twofold rotation, $C_{2a}$ in AA and $C_{2b}$ in AB, exchanges the two opposite-spin sublattices. Combined with $U_{\pi}$, these operations
enforce the sign-changing momentum-dependent spin splitting characteristic
of the $i$-wave altermagnetic state. The spin-space-group analysis and the corresponding symmetry generators are summarized in Table~\ref{tab:ssg_bilayers}.

\begin{table}[t]
\centering
\footnotesize
\caption{Summary of the spin-space-group (SSG) symmetry analysis, with the monolayer taken in its relaxed structure. For each system the first generator preserves the two spin sublattices and the second, combined with the spin flip $U_\pi$, exchanges them.}
\label{tab:ssg_bilayers}
\renewcommand{\arraystretch}{1.25}
\setlength{\tabcolsep}{4pt}
\begin{tabular}{l c l c}
\toprule
\textbf{System} & \begin{tabular}[c]{@{}c@{}}\textbf{Space}\\\textbf{group}\end{tabular} & \textbf{SSG decomposition} & \textbf{Generators}\tabularnewline
\midrule
Monolayer  & $P31m$       & $[E\Vert P3] + [U_\pi\Vert\,\sigma_v P3]$           & \begin{tabular}[c]{@{}c@{}}$[E\Vert C_{3z}]$\\$[U_\pi\Vert\,\sigma_v]$\end{tabular}\tabularnewline
AA bilayer & $P\bar{3}m1$ & $[E\Vert P\bar{3}] + [U_\pi\Vert\,C_{2a} P\bar{3}]$ & \begin{tabular}[c]{@{}c@{}}$[E\Vert IC_{3z}]$\\$[U_\pi\Vert\,C_{2a}]$\end{tabular}\tabularnewline
AB bilayer & $P321$       & $[E\Vert P3] + [U_\pi\Vert\,C_{2b} P3]$             & \begin{tabular}[c]{@{}c@{}}$[E\Vert C_{3z}]$\\$[U_\pi\Vert\,C_{2b}]$\end{tabular}\tabularnewline
\bottomrule
\end{tabular}
\end{table}

\subsection{Symmetry operations protecting the $i$-wave state}

In the non-SOC limit the Hamiltonian is spin-diagonal, with the splitting $\Delta E(\kq)=E_{\uparrow}(\kq)-E_{\downarrow}(\kq)$. Three operations fix the angular form of the splitting in all three systems.
\begin{itemize}
  \item \textbf{Spin-exchanging generators} $[\Upi\Vert R]$ with $R\in\{\sigma_v,C_{2a},C_{2b}\}$. They map $(\uparrow,\kq)\to(\downarrow,R\kq)$ and enforce $\Delta E(R\kq)=-\Delta E(\kq)$, so the splitting is odd under a reflection of $\kq$ and vanishes on the reflection line.
  \item \textbf{Spin-preserving} $C_{3z}\in\HH$. It enforces $\Delta E(C_{3z}\kq)=\Delta E(\kq)$, a period of $120^\circ$ in the polar angle $\phi$ of $\kq$.
  \item \textbf{Spinless time reversal} $S$, the product of time reversal and a spin rotation by $\pi$ about an axis perpendicular to the moments. It is a symmetry of every collinear magnet with real hoppings and negligible spin-orbit coupling, acts within each spin channel as $E_{\sigma}(\kq)=E_{\sigma}(-\kq)$, and hence gives $\Delta E(-\kq)=\Delta E(\kq)$, a period of $180^\circ$ in $\phi$.
\end{itemize}
The last two together give $\Delta E$ a period of $60^\circ$, because the only angular harmonics with both a $120^\circ$ and a $180^\circ$ period are $l=0,6,12,\dots$, and the first removes $l=0$ because an odd function has no constant part. Every harmonic below $l=6$ therefore vanishes identically. The lowest-order polynomial compatible with all three constraints is $k^{6}\cos 6\phi$, a form factor with an angular period of $\pi/3$ that changes sign every $\pi/6$ and has twelve nodal directions in the Brillouin zone, which is the $i$-wave ($l=6$) angular dependence~\cite{le2026catalog}.

\section{RPA spin response and spin stiffness}\label{app:rpa}
In the two-dimensional isotropic limit with negligible spin-orbit coupling, the Mermin-Wagner theorem rules out true long-range antiferromagnetic order at finite temperature, so a conventional N\'eel temperature is not well defined within these models~\cite{mermin1966absence}. Exchange interactions nevertheless sustain short-range antiferromagnetic correlations over a finite temperature range, and we therefore estimate the spin stiffness as the relevant magnetic energy scale.  

\begin{figure*}[t]
    \centering
    \includegraphics[width=1.\linewidth]{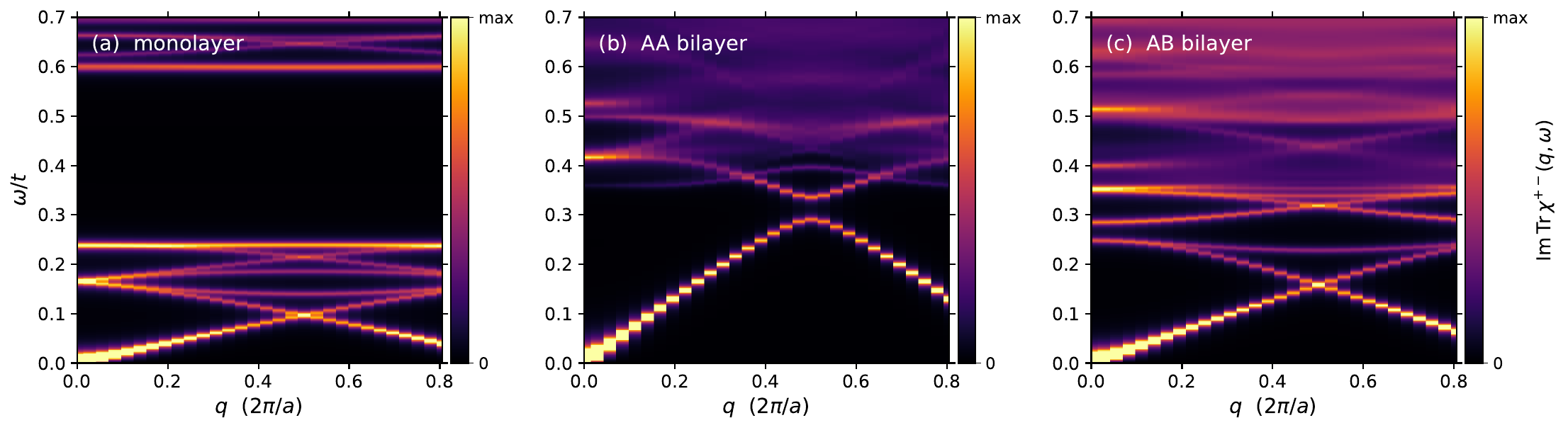}
    \caption{RPA transverse spin susceptibility $\chi''(q,\omega)$ for (a) the monolayer, (b) the AA-stacked bilayer, and (c) the AB-stacked bilayer graphene superlattices. The color scale is linear in $\mathrm{Im}\,\mathrm{Tr}\chi^{+-}$ and is clipped at the 99th percentile of each panel. The gapless branch emerging from $q=0$ is the acoustic magnon, whose slope gives the spin-wave velocity. All three panels share one grid, with $n_k=36$ and 30 momenta $q=n/36$.}
    \label{fig:PRA_J_1}
\end{figure*}

To estimate the characteristic magnetic energy scale, we calculate the transverse spin susceptibility within the random-phase approximation (RPA) and extract the low-energy acoustic magnon branch from the resulting spin excitation spectrum (Fig.~\ref{fig:PRA_J_1}). A gapless magnon branch appears at the $\Gamma$ point, as required by the Goldstone theorem~\cite{wagner1966long}. Near the $\Gamma$ point the magnon dispersion is linear,
\begin{equation}
\frac{\omega}{t} \simeq v q ,
\end{equation}
where $q$ is measured in units of the superlattice reciprocal vector $\mathbf{b}_1$. We obtain $v=0.198$, $0.647$, and $0.323$ for the monolayer, AA-stacked bilayer, and AB-stacked bilayer, respectively. For the monolayer, this corresponds to $\hbar v = 1.88~\mathrm{eV}\,\text{\AA}$. The velocity is a direct output of the calculation and requires no modeling.

The quantity that controls the ordering temperature is not an exchange constant but the spin stiffness $\rho_s$, the energy cost of a long-wavelength twist of the N\'eel vector. We obtain it by mapping the RPA magnon onto linear spin-wave theory on the real-space nearest-neighbor network of the superlattice,
\begin{equation}
\rho_s
=
\frac{J}{A_{\mathrm{cell}}}
\sum_{\langle ij\rangle}
S_i S_j \, (\hat{q}\cdot\boldsymbol{\delta}_{ij})^2 ,
\end{equation}
where $S_i$ is the converged local moment on site $i$, $\boldsymbol{\delta}_{ij}$ is the bond vector, $A_{\mathrm{cell}}$ is the area of the magnetic unit cell, and $J$ is the overall energy scale of the mapping, fixed by matching the Heisenberg magnon slope to the RPA velocity. We evaluate the stiffness with the computed site-resolved moments rather than with a nominal $S=1/2$.

This gives spin stiffnesses of $\rho_s \approx 177~\mathrm{meV}$ for the monolayer, $\rho_s \approx 158~\mathrm{meV}$ for the AA-stacked bilayer and $\rho_s \approx 159~\mathrm{meV}$ for the AB-stacked bilayer. Although the magnon velocities differ substantially, the stiffnesses remain comparable because the steeper branches of the bilayers are compensated by their smaller local magnetic moments. All values are approximately six to seven times larger than the room-temperature thermal energy scale of $k_{\mathrm B}T \approx 25$~meV. Hence, the antiferromagnetic correlation length stays large at room temperature even though the isotropic two-dimensional limit forbids true long-range order. Since the magnetization in the vacancy superlattices is spatially nonuniform, the stiffness rather than a bond-resolved exchange constant is the meaningful characterization of these systems.

\section{Local-moment formation temperature}\label{app:tloc}

\begin{figure}[t]
    \centering
    \includegraphics[width=\columnwidth]{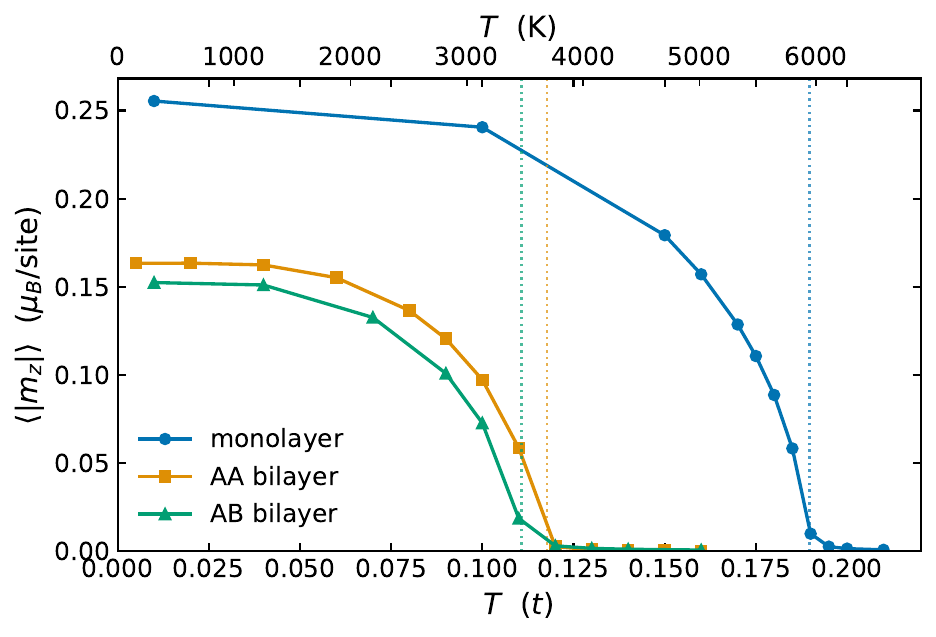}
    \caption{Site-averaged local moment $\langle |m_z| \rangle$ as a function of temperature, from finite-temperature mean-field theory at $U=2t$. The moments vanish at $T_{\mathrm{loc}}$ (dotted lines), obtained by extrapolating $\langle |m_z| \rangle^2$ to zero. This is the temperature at which the local moments cease to exist, and not a N\'eel temperature, since long-range order is absent at any finite temperature in the isotropic two-dimensional limit.}
    \label{fig:mag_vs_T}
\end{figure}

To complement the RPA spin-stiffness analysis we have performed finite-temperature mean-field calculations for the three structures, raising the temperature in the self-consistent Hubbard mean field until the magnetic solution collapses. The order parameter is the site-averaged local moment $\langle |m_z| \rangle$. The net moment vanishes at all temperatures because the state is a compensated antiferromagnet. The collapse temperature $T_{\mathrm{loc}}$ is obtained by extrapolating $\langle |m_z| \rangle^2$ to zero, which is linear in $T$ near the transition within mean-field theory. This gives $k_B T_{\mathrm{loc}} = 512$, $318$ and $299$~meV, that is $5940$, $3690$ and $3470$~K, for the monolayer, the AA-stacked bilayer and the AB-stacked bilayer (Fig.~\ref{fig:mag_vs_T}).

The two temperature scales measure different things. The mean-field calculation locates the temperature at which the local moments themselves disappear. Above $T_{\mathrm{loc}}$ the self-consistent solution no longer supports a finite $\langle |m_z| \rangle$ on any site, so there is no magnetism left to order. It is a local-moment formation scale, not an ordering temperature, and mean-field theory overestimates it because it omits the fluctuations that dominate a two-dimensional magnet. The onset of long-range order is instead controlled by the spin stiffness extracted from the RPA magnon.

The Mermin-Wagner theorem forbids long-range order only in the strictly isotropic limit. Spin-orbit coupling in these carbon systems is weak but finite, and the resulting magnetic anisotropy opens a gap in the magnon spectrum which cuts off the logarithmic infrared divergence of the thermal magnon population that would otherwise destroy the order. Because that cutoff enters only inside a logarithm, the ordering temperature is set by the spin stiffness and is only weakly sensitive to the size of the anisotropy, so even the small values computed here, $0.017$~meV per C site for the monolayer and $0.006$~meV per C site for the bilayers, are sufficient to stabilize order at a finite temperature. This is the same mechanism that gives monolayer CrI$_3$ a Curie temperature of $45$~K despite its two-dimensionality~\cite{lado2017origin}.

This allows a direct estimate. Balancing the thermal magnon population against the ordered moment gives an ordering temperature of the form
\begin{equation}
k_B T_N \;\sim\; \frac{\rho_s}{\ln\!\left(\rho_s/\Delta\right)},
\label{eq:tn}
\end{equation}
with $\Delta$ the anisotropy gap and a prefactor of order unity. Taking the measured intrinsic spin-orbit gap of graphene, $\Delta \approx 42~\mu$eV~\cite{sichau2019resonance}, together with the stiffnesses obtained above gives $\ln(\rho_s/\Delta) \approx 8$, and hence $k_B T_N \approx 21$~meV ($245$~K) for the monolayer and $k_B T_N \approx 19$~meV ($225$~K) for the AA- and AB-stacked bilayers.
The estimate is only weakly sensitive to the anisotropy, which is its least certain input. Replacing $\Delta = 42~\mu$eV by the anisotropies computed here from first principles, $17~\mu$eV and $6~\mu$eV per C site, changes the logarithm by at most $24\%$ and moves $T_N$ to $220$ and $180$~K. A sevenfold uncertainty in the anisotropy therefore costs less than a factor $1.3$ in the ordering temperature, because the anisotropy enters only inside the logarithm while the stiffness sets the scale. These systems therefore order at a temperature of the order of room temperature, although no particular value should be taken as exact.

The mean field thus bounds the magnetic energy scale from above and the RPA stiffness sets the ordering scale below it, the former twelve to twenty times and the latter six to seven times the room-temperature thermal energy.

\section{Magnonic detection of the $i$-wave order}\label{app:magnon}

The graphene superlattices studied here are magnetic insulators, with a tight-binding gap of $1.96$~eV for the monolayer and $0.99$~eV for both bilayers and the chemical potential in the gap. This rules out the charge-transport fingerprints usually invoked for altermagnets. With the chemical potential in the gap the occupation factor is unity on every valence band and zero on every conduction band, so any Drude-type response reduces to the Brillouin-zone integral of a momentum derivative of a smooth periodic function, which vanishes identically at every order in the driving field. The absence of a low-order transport response is then a property shared by any band insulator and carries no information about the wave symmetry.

There is a further, purely symmetry-based obstruction. An $i$-wave form factor is an $l=6$ angular harmonic, whereas a response tensor of rank $r$ carries angular components only up to $m=r$. Any Brillouin-zone-integrated response of rank five or lower is therefore blind to $i$-wave order regardless of the filling. Momentum-resolved probes are not subject to this constraint, and two such routes are described in the following.

First, spin-resolved angle-resolved photoemission spectroscopy (spin-ARPES) provides a direct probe of the electronic spin splitting in occupied bands lying several eV below the Fermi level~\cite{ngabonziza2025layer}. It reconstructs the spin-resolved band dispersion from the energy, emission angle, and spin polarization of photoemitted electrons. Spin-resolved photoemission has resolved spin splittings of about 100 meV in the Dirac cone of graphene on Au~\cite{marchenko2012giant} and of up to 0.5 eV in WSe$_2$~\cite{riley2014direct}, and photoemission with a micrometer-scale focus now maps the band structure of exfoliated flakes and devices, including twisted bilayer graphene~\cite{lisi2021observation}, encapsulated few-layer WTe$_2$~\cite{cucchi2019microfocus} and gated graphene heterostructures~\cite{nguyen2019visualizing}. The pronounced higher-energy splitting predicted in our systems should therefore be directly accessible by spin-ARPES, as demonstrated recently for altermagnetic MnTe~\cite{krempasky2024altermagnetic} and CrSb~\cite{yang2025three}.

Second, the magnon spectrum is the standard probe of altermagnetism in insulators. In a compensated collinear magnet the two magnon branches carry opposite spin chirality, and in an altermagnet they are split by an amount that follows the same angular form factor as the electronic spin splitting,
\begin{equation}
\omega_{+}(\kq) - \omega_{-}(\kq) \;\propto\; \Delta(\kq) .
\label{eq:magnonsplitSI}
\end{equation}
The magnon splitting therefore inherits the nodal structure of the band splitting. For the $i$-wave state realized here this gives a sixfold-periodic, sign-alternating magnon splitting with twelve nodal directions in $\kq$, the same twelve directions identified in the electronic structure, now appearing in the magnon dispersion. Because the effect is nonrelativistic it does not require spin-orbit coupling, which is an advantage in a carbon-based platform, where a measured splitting cannot be attributed to relativistic effects.

\begin{figure}[t]
    \centering
    \includegraphics[width=\columnwidth]{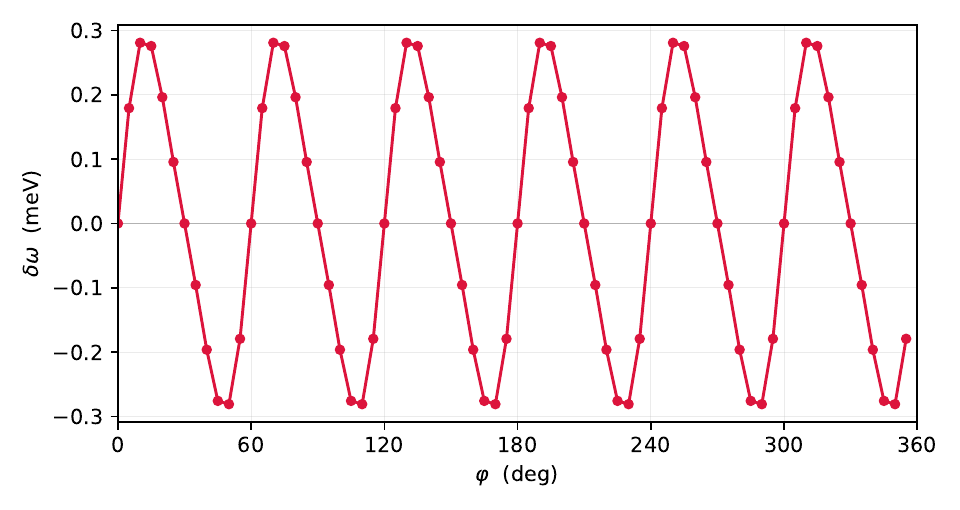}
    \caption{Chiral magnon splitting on a circular path of fixed $|q|$ in the AA-stacked bilayer. The plotted quantity is the part of the magnon energy that is odd under the mirror $\varphi\to60^\circ-\varphi$, namely $\delta\omega(\varphi)=[\omega(\varphi)-\omega(60^\circ-\varphi)]/2$. That mirror exchanges the two magnetic sublattices, so the nonmagnetic hexagonal warping of the lattice, which is $25.3$~meV peak to peak and would otherwise bury the effect, is even under it and cancels exactly. Every mirror partner is a computed point of the $5^\circ$ grid, so nothing is interpolated. Obtained from the poles of the RPA kernel at $n_k=36$, with $|q|=0.55\times2\pi/a$.}
    \label{fig:magnon_angle}
\end{figure}

We have computed this splitting directly. Evaluating the poles of the RPA kernel on a circle of fixed $|q|$ in the AA-stacked bilayer gives the magnon energy $\omega(\varphi)$ around the circle. That energy is dominated by the nonmagnetic hexagonal warping of the lattice, which is $25.3$~meV peak to peak and would bury the effect of interest. The altermagnetic contribution is instead the part of $\omega(\varphi)$ that is odd under the mirror $\varphi\to60^\circ-\varphi$. That mirror exchanges the two magnetic sublattices and is broken only by the magnetic order, whereas the warping is even under it and cancels exactly. This odd part is shown in Fig.~\ref{fig:magnon_angle}. It is sixfold periodic with twelve nodes, carries 97\% of its angular power in the $l=6$ harmonic, and splits the two magnon branches by $0.56$~meV. The agreement of that harmonic content with the electronic form factor confirms that the magnon splitting carries the same $i$-wave symmetry as the band splitting.

The radial path used for the $\chi''$ maps of Fig.~\ref{fig:PRA_J_1} runs along $\mathbf{b}_1$, which is one of these twelve nodal directions, so those maps show no chiral splitting at any level of convergence. A circular path is needed to see it, which is why the splitting is presented as a function of angle rather than as a further panel of Fig.~\ref{fig:PRA_J_1}.

This signature is established in comparable materials. Chiral magnon splitting with alternating chirality was predicted from first principles for the $d$-wave altermagnet RuO$_2$~\cite{smejkal2023chiral} and has been identified in bulk altermagnetic MnTe~\cite{chiral2024mnte}. In a two-dimensional sample the magnon spectrum is reached through magnon spin transport. Long-distance magnon transport has been measured in atomically thin van der Waals antiferromagnets~\cite{xing2019magnon}, and the spin-split magnon bands of an insulating altermagnet carry spin Seebeck and spin Nernst currents that require neither a magnetic field nor spin-orbit coupling~\cite{cui2023efficient}, as shown for a collinear honeycomb geometry closely related to ours~\cite{sarkar2025magnon}. Noise magnetometry has been proposed as a way of separating altermagnetic from antiferromagnetic order through the magnon contribution alone, with no mobile carriers required~\cite{pupim2026noise}. The electronic spin splitting that sets the magnon form factor is in turn directly accessible to spin-ARPES, as described above.

The magnon splitting is robust against averaging over time-reversed altermagnetic domains, provided the domains are large compared with the length scale set by the momentum resolution of the probe. Each domain hosts the two branches at $\omega_0\pm\Delta(\kq)$, and the time-reversed domain only exchanges which branch carries which spin chirality, so a domain average retains the splitting and its twelve nodal directions. What averaging cancels is the chirality-resolved sign of the splitting, together with the time-reversal-odd magnon spin Nernst and Seebeck signals, and recovering that sign requires a single-domain measurement or an unequal population of the two domains.

\section{Supercell selection and symmetry preservation}\label{app:supercell}

The tight-binding calculations employ a $6\sqrt{3}\times6\sqrt{3}$ supercell for the monolayer, a $3\sqrt{3}\times3\sqrt{3}$ supercell for the AA-stacked bilayer and a $5\sqrt{3}\times5\sqrt{3}$ supercell for the AB-stacked bilayer. The first-principles calculations use $6\times6$, $3\sqrt{3}\times3\sqrt{3}$, and $5\times5$ supercells, respectively. The tight-binding cells follow the antidot superlattice construction, while for the first-principles calculations we adopt the smallest computationally affordable cells that preserve the same target spin point-group symmetry as the corresponding tight-binding models. We have verified that these first-principles structures retain the required symmetry and exhibit the same $i$-wave altermagnetic spin-splitting pattern. Although the tight-binding and first-principles calculations therefore correspond to different superlattice periodicities for the monolayer and the AB-stacked bilayer, they realize the same altermagnetic symmetry.

\section{Feasibility of AA-stacked graphene}\label{app:aa}
AA stacking is not the ground-state configuration of pristine bilayer graphene, but AA-stacked bilayer graphene is experimentally accessible. AA-stacked phases were identified in epitaxially grown bilayer graphene on Ru(0001)~\cite{papagno2012two}. In addition, AA-stacked regions are directly imaged by scanning tunneling microscopy in marginally twisted bilayer graphene, where they carry a pronounced localized tunneling peak and can be tuned by strain~\cite{ouyang2026structural}. A local AA environment is thus accessible even without globally AA-stacked material.

\clearpage
\bibliography{biblio_clean}

\end{document}